\documentclass[final,3p,times,authoryear]{elsarticle}
\usepackage[T1]{fontenc}
\usepackage[utf8]{inputenc}
\usepackage{graphicx}
\usepackage{amssymb}
\usepackage{amsmath}
\usepackage{color}
\usepackage{siunitx}
\usepackage{subcaption}
\usepackage{booktabs}
\usepackage{hyperref}
\usepackage{cleveref}
\usepackage{pgfplots}
\usepackage{xspace}
\usepackage[version=4]{mhchem}

\usepackage[makeindex]{imakeidx}
\usepackage{datatool}
\newcommand*{\addabbr}[2]{%
  \DTLnewrow{abbreviations}%
  \DTLnewdbentry{abbreviations}{Abbreviation}{#1}%
  \DTLnewdbentry{abbreviations}{Description}{#2}%
}
\newcommand*{\addsymb}[2]{%
  \DTLnewrow{symbols}%
  \DTLnewdbentry{symbols}{Symbol}{#1}%
  \DTLnewdbentry{symbols}{Description}{#2}%
}
\newenvironment{nomenclature}{\begin{list}{}%
    {%
    \setlength{\labelwidth}{3cm}\setlength{\leftmargin}%
    {\labelwidth+\labelsep}%
\setlength{\itemsep}{0pt}}}{\end{list}}

\graphicspath{{./figures/}{./tikz/}}
\usetikzlibrary{arrows.meta}
\usepgfplotslibrary{external}
\usepgfplotslibrary{colorbrewer}
\usepgfplotslibrary{groupplots}
\usepgfplotslibrary{fillbetween}
\usepgfplotslibrary{statistics}
\pgfplotsset{
  compat=newest,
  every axis/.append style={
    height=0.8\textwidth,
    width=1.0\textwidth,
    ylabel style={at={(-0.12,0.5)}},
    axis lines=left,
    axis line style={thick, -{Latex[length=5pt,width=5pt]}},
    every tick/.style={color=gray},
    legend style = {
      draw=none,
      fill=black!3,
      font=\footnotesize,
    },
    legend cell align = {left},
    legend pos = north east,
    legend image code/.code={
      \draw[mark repeat=2,mark phase=2]
      plot coordinates {
        (0cm, 0cm)
        (0.2cm, 0cm)
        (0.4cm, 0cm)
      };
    },
    font=\small,
    no markers,
    samples=100,
    every axis plot/.append style={thick},
    set layers=axis on top,
  },
  cl3-1/.style={color=YlGnBu-G},
  cl3-2/.style={color=YlGnBu-E},
  cl3-3/.style={color=YlGnBu-D},
  cl4-1/.style={color=YlGnBu-I},
  cl4-2/.style={color=YlGnBu-G},
  cl4-3/.style={color=YlGnBu-E},
  cl4-4/.style={color=YlGnBu-D},
  emph/.style={
    very thick,
    densely dashed,
  },
  sim/.style={only marks},
  table/col sep = comma,
  table/x = x,
  /pgf/declare function={
    g(\rho) = 9.81*(1.0 - \rho/1000.0);
    u(\S,\rho,\rnull) = pow(pow(27, 1/2)*g(\rho)*\S/(2*pi*\rho*\rnull), 1/3);
    r_rpt(\theta,\S,\dHqdot) = sqrt(\theta*\S*\dHqdot/pi);
    t_rpt(\theta,\S,\dHqdot,\rho,\rnull)
      = r_rpt(\theta,\S,\dHqdot)/u(\S,\rho,\rnull)
          *(1 + sqrt(2))
          *(1 - 4*sqrt(\rnull/(3*sqrt(3)*r_rpt(\theta,\S,\dHqdot))) + 5*ln(r_rpt(\theta,\S,\dHqdot)/\rnull)*\rnull/(3*r_rpt(\theta,\S,\dHqdot)*sqrt(3)));
    m_rpt(\t,\theta,\S,\dHqdot,\rho,\rnull)
      = \S*(1-\theta)*(\t - t_rpt(\theta,\S,\dHqdot,\rho,\rnull));
  },
}

\usepackage{todonotes}
\presetkeys{todonotes}{
  inline,
  bordercolor=gray!60,
  backgroundcolor=gray!60,
}{}
\makeatletter
\renewcommand{\todo}[2][]{%
  \tikzexternaldisable\@todo[#1]{#2}\tikzexternalenable%
}
\makeatother

\journal{Journal of Loss Prevention in the Process Industries}
\newcommand*{\papertitle}{A combined fluid-dynamic and thermodynamic model to predict the onset of rapid phase transitions in LNG spills}
\newcommand*{\paperkeywords}{%
  Thermodynamics\sep
  Equation of State\sep
  Spinodal\sep
  Nucleation\sep
  Phase Stability\sep
  Rapid Phase Transition\sep
  Liquid Spills\sep
  Risk Assessment\sep
  Explosions
}

\hypersetup{
  pdfauthor = {Karl Yngve Lervåg, Hans L. Skarsvåg, Jabir Ali Ouassou, Eskil Aursand, Morten Hammer, Gunhild Reigstad, Øivind Wilhelmsen, Eirik Holm Fyhn, Magnus Aa. Gjennestad, Peder Aursand, Åsmund Ervik},
  pdftitle = {\papertitle},
  pdfsubject = {LNG spills and consequent RPT},
  pdfkeywords = {\paperkeywords},
  pdfcreator = {LaTeX with hyperref package},
  bookmarksnumbered,
  pdfstartview={FitH},
  linkcolor=black,
  colorlinks=true,
}

\DeclareSIUnit{\masspercent}{wt\%}

\crefname{equation}{Eq.}{Eqs.}
\crefname{section}{Section}{Sections}
\crefname{table}{Table}{Tables.}
\crefname{figure}{Figure}{Figures}
\crefname{subfigure}{Figure}{Figures.}

\let\OLDitemize\itemize
\renewcommand\itemize{\OLDitemize\setlength{\itemsep}{0em}}

\newcommand*{\water}{\mathrm{w}}
\newcommand*{\liq}{\mathrm{l}}
\newcommand*{\vap}{\mathrm{v}}

\newcommand*{\shl}{\mathrm{SHL}}
\newcommand*{\leid}{\mathrm{L}}
\newcommand*{\crit}{\mathrm{c}}

\newcommand*{\vapfilm}{\mathrm{vf}}
\newcommand*{\evap}{\mathrm{evap}}
\newcommand*{\eff}{\mathrm{eff}}

\newcommand*{\spill}{\mathrm{spill}}
\newcommand*{\release}{\mathrm{release}}
\newcommand*{\bub}{\mathrm{bub}}
\newcommand*{\atm}{\mathrm{atm}}

\newcommand*{\LNG}{\mathrm{LNG}}
\newcommand*{\LE}{\mathrm{LE}}

\newcommand*{\RPT}{\mathrm{RPT}}
\newcommand*{\capillary}{\mathrm c}
\newcommand*{\spinodal}{\mathrm{spinodal}\xspace}
\newcommand*{\eos}{\ensuremath{\text{EoS}}\xspace}

\newcommand*{\Ra}{\mathrm{Ra}}

\newcommand*{\pdd}[2]{\ensuremath{\frac{\partial #1}{\partial #2}}}
\newcommand*{\pd}[1]{\pdd{}{#1}}
\newcommand*{\dd}{\ensuremath{\mathop{}\!\mathrm{d}}}
\newcommand*{\pdtot}[2]{\ensuremath{\frac{\dd #1}{\dd #2}}}
\newcommand*{\pdtoti}[2]{\ensuremath{{\dd #1}/{\dd #2}}}
\newcommand{\vect}[1]{\boldsymbol{#1}}
\renewcommand{\div}{\ensuremath{\nabla\cdot}}
\newcommand*{\abs}[1]{\ensuremath{\left\lvert #1 \right\rvert}}

\newcommand*{\eg}{e.g.~}
\newcommand*{\ie}{i.e.~}

\begin{document}

\begin{frontmatter}
  \title{\papertitle}
  \date{\today}

  \author[sintef]{Karl Yngve Lervåg\corref{cor1}}
  \ead{karl.lervag@sintef.no}
  \author[sintef]{Hans~L.~Skarsvåg}
  \author[sintef,ntnu]{Eskil~Aursand}
  \author[sintef]{Jabir~Ali~Ouassou}
  \author[sintef]{Morten~Hammer}
  \author[sintef]{Gunhild~Reigstad}
  \author[sintef]{Åsmund~Ervik}
  \author[sintef,ntnu2]{Eirik~Holm~Fyhn}
  \author[sintef]{Magnus~Aa.~Gjennestad}
  \author[sintef]{Peder~Aursand}
  \author[sintef,ntnu]{Øivind~Wilhelmsen}
  \address[sintef]{SINTEF Energy Research, P.O. Box 4671 Sluppen, NO-7465 Trondheim, Norway}
  \address[ntnu]{NTNU, Department of Energy and Process Engineering, Kolbjørn Hejes v 1B, NO-7491 Trondheim, Norway}
  \address[ntnu2]{NTNU, Center for Quantum Spintronics, Department of Physics, Høgskoleringen 5, NO-7491 Trondheim, Norway}

  \begin{abstract}
  Transport of liquefied natural gas (LNG) by ship occurs globally on a massive scale.
  The large temperature difference between LNG and water means LNG will boil violently if spilled onto water.
  This may cause a physical explosion known as rapid phase transition (RPT).
  Since RPT results from a complex interplay between physical phenomena on several scales, the risk of its occurrence is difficult to estimate.
  In this work, we present a combined fluid-dynamic and thermodynamic model to predict the onset of delayed RPT.
  On the basis of the full coupled model, we derive analytical solutions for the location and time of delayed RPT in an axisymmetric steady-state spill of LNG onto water.
  These equations are shown to be accurate when compared to simulation results for a range of relevant parameters.
  The relative discrepancy between the analytic solutions and predictions from the full coupled model is within \SI{2}{\percent} for the RPT position and within \SI{8}{\percent} for the time of RPT.
  This provides a simple procedure to quantify the risk of occurrence for delayed RPT for LNG on water.
  Due to its modular formulation, the full coupled model can straightforwardly be extended to study RPT in other systems.
  \end{abstract}

\end{frontmatter}

\section{Introduction}
\label{sec:introduction}
Natural gas is a common fossil fuel used for heating, cooking, propulsion and electricity-generation across the globe.
Its main component is methane (about \SI{90}{\masspercent}), with the remainder consisting of progressively smaller amounts of the heavier alkanes, such as ethane, propane, and butane.
For the purpose of long-range transportation, natural gas is cooled in large refrigeration systems to form liquefied natural gas (LNG)~\citep{kumar2011}.
LNG is transported across the world's oceans in large carriers, and a single carrier may carry up to around \SI{260000}{\meter\cubed} of LNG. Furthermore, the interest of using LNG as fuel in shipping to reduce CO$_2$ emissions is increasing~\citep{schinas2016feasibility,lee2015fire}. Combined with an increasing trend towards both processing and usage of LNG at sea~\citep{dnv-lng-fuel-report1}, this means that there are many scenarios where LNG may be accidentally spilled and come in contact with seawater.

\begin{figure}[tbp]
  \centering
  \includegraphics[width=0.65\textwidth]{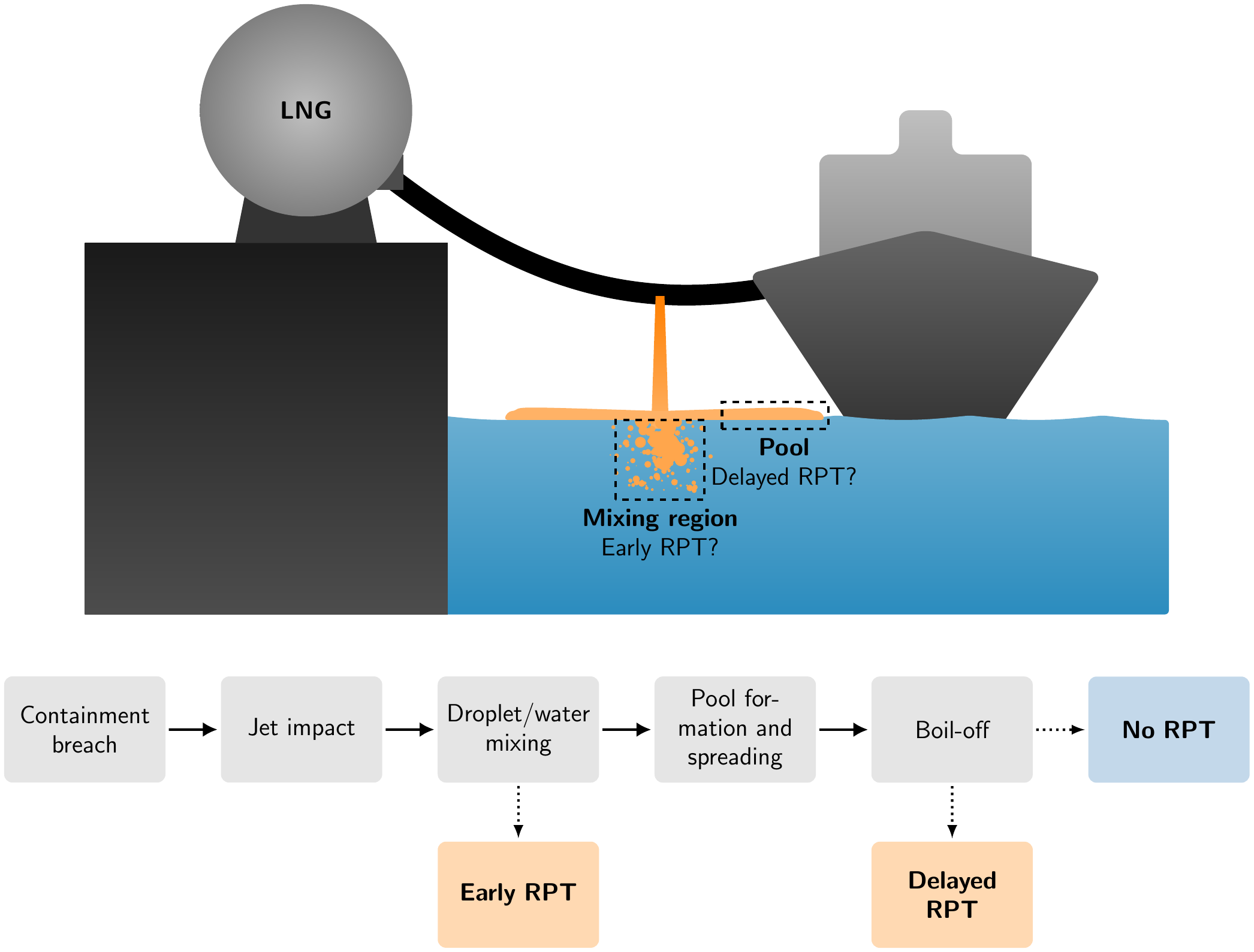}
  \caption{An illustration of an accidental spill event during loading or unloading of LNG.
  The flow chart shows the possible pathways of the different kinds of RPT events.
  The jet from a containment breach impacts the water and forms a mixing region beneath the surface.
  At this location, there is a known possibility of \emph{early RPT}.
  Since the LNG density is lower than that of water, the LNG droplets rise and form a floating pool.
  This pool then spreads out while simultaneously changing its composition via boil-off.
  Eventually, this change may result in a \emph{delayed RPT}.
  In this work, we focus on the delayed RPT.}
  \label{fig:spill-event}
\end{figure}

The chain of events of a marine LNG spill consists of several steps as illustrated in \cref{fig:spill-event}.
Initially, there must be a \emph{containment breach} where the containment of LNG in a tank or transfer line is broken.
If the loss of containment (LOC) is above sea level, the LNG may fall towards the water surface.
When the LNG jet impacts the water surface, it would break up.
If the momentum of the jet is large enough, the droplets initially penetrate the surface and become submerged in water, which forms a chaotic \emph{mixing region}.
Since the density of LNG is about half that of water, the droplets are buoyant and rise to the surface.
Subsequently, the spill forms a pool that spreads on top of the water surface.
The boiling point of LNG is about \SI{-162}{\celsius}, so the pool starts to boil while spreading.
Since methane is the most volatile component, the resulting vapor is almost pure methane.
This causes a gradual compositional change, which increases the relative amounts of the heavier alkanes such as ethane, propane and butane.

In most cases, the spill proceeds to complete boil-off.
However, in some cases it is observed to suddenly, and seemingly at random, undergo a localized explosive vaporization.
This phenomenon was discovered in the 1960s and is called a rapid phase transition (RPT), although sometimes referred to as a \emph{vapor explosion} or \emph{physical explosion}.
This is not to be confused with a combustive chemical explosion, but it is still destructive in nature and poses a hazard to both people and equipment.
Its peak pressures and released mechanical energy can be large enough to displace and damage heavy equipment~\citep{luketa2006,pitblado2011,forte2017} and could cause secondary structural damage and cascading containment failures~\citep{havens2007}.
The yields of single RPT events vary greatly, and may have TNT equivalents ranging from a few grams to several kilograms, which corresponds to about \SIrange{0.01}{25}{\mega\joule}~\citep{koopman2007,iomosaic2006,cleaver1998,ABS2004,hightower2004}.
As such, RPT is considered one of the main safety concerns in the LNG industry~\citep{reid1983,pitblado2011}.
Still, the attention given to RPT risk and hazard in LNG safety reviews is highly variable, ranging from significant discussions~\citep{pitblado2011,cleaver2007,luketa2006,shaw2005} to little more than a brief mention~\citep{alderman2005,hightower2005,havens2007,raj2010,forte2017}.
There may also be a risk of RPT occurring during transport of other cryogenic liquids such as liquid hydrogen~\citep{wilhelmsen2018,liu2019}, argon, helium, nitrogen, and oxygen~\citep{reid1983}.
There is an increasing focus on hydrogen as a clean energy carrier, and liquid hydrogen at about \SI{-250}{\celsius} has been described as a promising mode for large-scale transport~\citep{wilhelmsen2018}.
This merits further development of models and theoretical predictions for the occurrence of RPT in cryogenic fluids.

The large temperature difference between LNG and water suggests that the interaction of these fluids may lead to ice formation in the water.
However, this has not been observed when LNG is spilled on open bodies of water~\citep{conrado2000}.
This can be explained by a circulation of the water sufficient to compensate for the heat transfer.

As illustrated in \cref{fig:spill-event}, there is an established distinction between two kinds of RPT events depending on when and where it occurs~\citep{luketa2006,koopman2007}:
\emph{Early RPT}, defined as any RPT that occurs in the mixing region at any time during the spill event; and \emph{delayed RPT}, defined as any RPT that is not an early RPT.
Experiments indicate that delayed RPT only occurs a considerable time after the start of the LNG spill event (on the scale of minutes)~\citep{iomosaic2006}.
The figure also illustrates the pathways along the previously discussed chain of events for the two different kinds of RPT.

In the first half of the 1970s, researchers arrived at a general consensus for a theory of RPT~\citep{katz1971,katz1972,nakanishi1971,enger1972,enger1972b,enger1973} which is described in depth by \citet[Ch.~1]{aursand2019}.
The main idea is that the LNG pool on water is initially within the \emph{film-boiling regime}.
In this case, the LNG pool is insulated from the water by a vapor film that keeps the heat flux low.
When the LNG composition changes due to boil-off of the volatile components, mostly methane, the film collapses due to a change of boiling regime~\citep{aursand2018a}.
The temperature at which this change occurs is known as the Leidenfrost temperature.
This leads to direct contact between water and LNG and a rapid increase in the heat flux.
The LNG pool may then be superheated and transition through a meta-stable state until the liquid approaches the \emph{superheat limit}~\citep{aursand2017}.
At the superheat limit, the LNG will spontaneously vaporize by \emph{homogeneous nucleation}.
This is called a rapid phase transition and will create a pressure wave and potentially release considerable energy through expansion work.

Whether or not an RPT event will occur in any given spill has been notoriously difficult to predict.
From extensive tests performed by Lawrence Livermore National Laboratory (LLNL) in the 1980s~\citep{luketa2006,koopman2007,iomosaic2006}, it was found that RPT occurred in about one third of all spills.
It was also observed that a single spill may lead to more than ten distinct RPT events.

\citet{cleaver2007} give a summary of the experimental data on LNG safety.
A more thorough history of LNG RPT research and experimental measurements is given by \citet[Ch.~1]{aursand2019}.
However, although there have been extensive testing, there is still a shortage of available data on RPT events for LNG on water.
In particular, there is a lack of data that is suitable as a reference for advanced models, especially with regard to RPT events~\citep{thyer2003}.

The main challenge when predicting the occurrence of RPT is to predict the sudden film-boiling collapse.
We will refer to this as the \emph{triggering event}.
The approach depends on whether one considers early RPT (droplet boiling) or delayed RPT (pool boiling).
The scope of the present paper is limited to delayed RPT.
In particular, we deal with the modelling and prediction of delayed RPT events as a consequence of some containment breach.

A comprehensive review of earlier work on modelling LNG spills is provided by \citet{webber2009}.
Another detailed overview is provided by \citet{hissong2007}.
The main focus of earlier works has been to model the spreading, evaporation, and dispersion.
To our knowledge, there has been little efforts towards predicting the onset of delayed RPT events as a consequence of spills.
One exception is \citet{horvat2018}, who presents a CFD methodology based on the homogeneous multiphase formulation for simulation of LNG spills and RPT.
He uses this methodology to simulate the behaviour of an LNG spill from point of release, with spreading, until the LNG evaporates and disperses.
He applies the Leidenfrost criterion to predict RPT events, but there is no further discussion or characterisation of the RPT events in particular.
Instead, the main focus is to predict the evaporated LNG cloud dispersion and flammability.

In our previous work~\citep{aursand2018b}, we proposed a method to determine how much boil-off of methane that is necessary to trigger RPT.
However, the method does not quantify \emph{when} and \emph{where} this criterion is reached.
This requires a more elaborate simulation of the spill event, which is the purpose of the present work.
In particular, a complete model for studying delayed RPT must include pool formation and spreading, heat transfer between the water and the LNG, and evaporation of LNG.
A thermodynamical equation of state (\eos) is necessary to describe the thermophysical properties and solving the multicomponent liquid--vapor phase transitions.

In this work, we present a multiscale modelling approach for predicting delayed RPT and analytic expressions for the time and location of delayed RPT.
The multiscale model relies on the shallow-water equations to capture how LNG spreads on water and includes models for evaporation, heat-transfer, and a state-of-the-art \eos.
The Leidenfrost criterion is used to predict the onset of RPT.
The analytic expressions are derived from an analysis of a continuous tank spill.
The methodology for predicting the RPT time and position is general and applicable to any steady-state scenario.
The main novelty of the work is both i) the multiscale modelling approach that couples the models for underlying phenomena for predicting the onset of RPT, and ii) the new analytic expressions for the time and location of delayed RPT.

The multiscale model and the numerical method used to solve it are presented in \cref{sec:spill_model}.
In \cref{sec:analytical}, we present an analysis of a continuous tank spill and derive simple, predictive models for the position and time of delayed RPT for this particular case.
A two-dimensional axisymmetric spill is analysed in detail and explicit estimates for RPT position and time are derived.
Next, in \cref{sec:contspill}, we consider the axisymmetric spill case with constant spill rate and show results from simulations.
The results are compared to the predictions from the analytical solutions derived in \cref{sec:analytical}.
The results show that the analytical models accurately reproduce the predictions from the more advanced simulation model.
Finally, in \cref{sec:conclusion}, we summarize the results and provide an outlook of possible future work.

\section{LNG spill model}
\label{sec:spill_model}
\begin{figure}[tbp]
  \centering
  \includegraphics[width=0.95\textwidth]{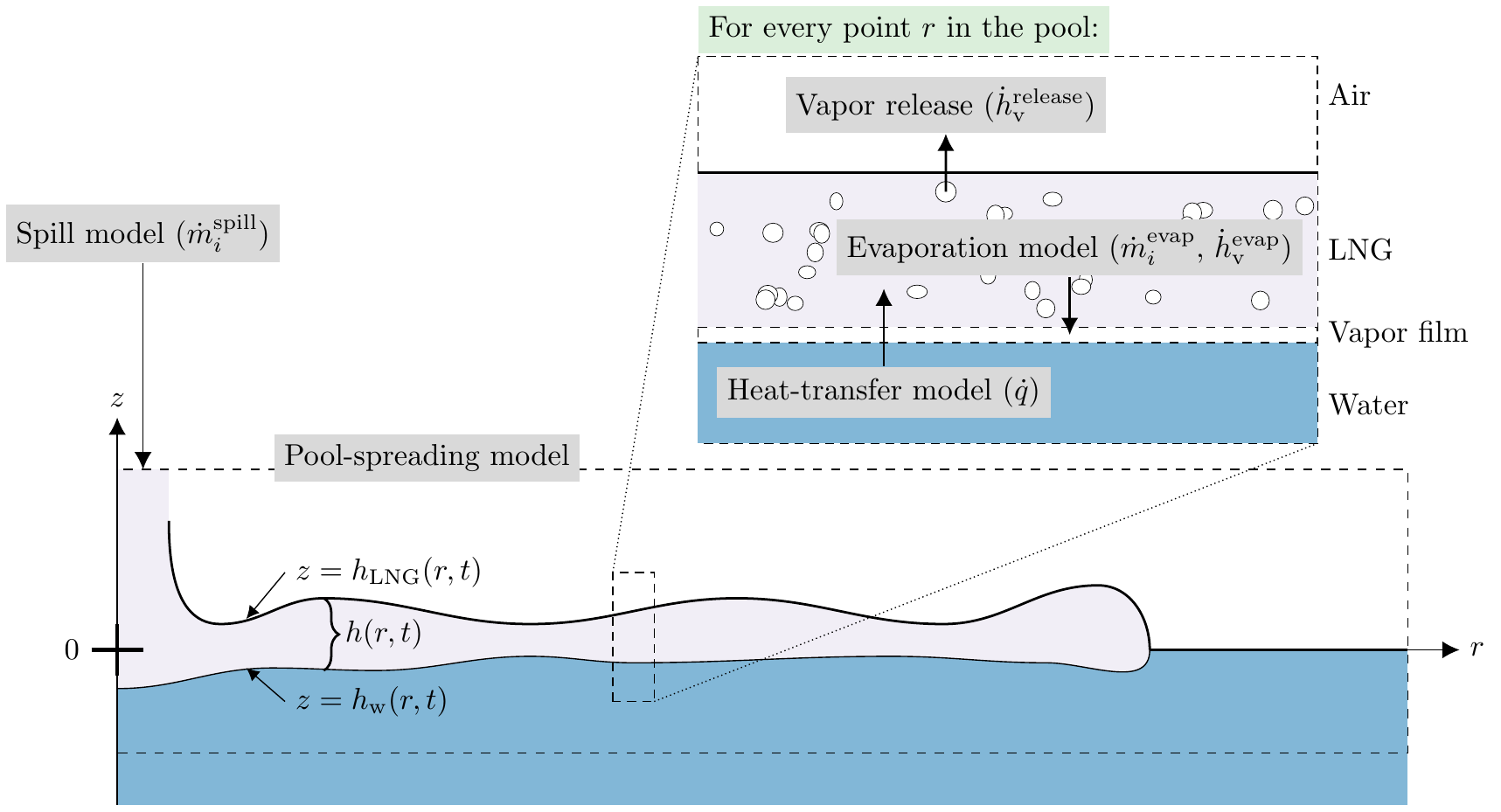}
  \caption{Illustration of the model considered in this work.
    The \textbf{pool-spreading model} handles the evolution of the LNG pool,
    which involves tracking its thickness~$h$ and the local thermodynamic state
    as functions of radius~$r$ and time~$t$.
    The vertical positions~$z$ of the interfaces are given by $h_\water$ and $h_\LNG$,
    which may be calculated from $h$ using \cref{eq:hwater,eq:hLNG}.
    The \textbf{spill model} calculates the added mass due to the spill,
    and the \textbf{heat transfer and evaporation models} calculate mass loss due to evaporation.
    These effects are added as source terms to the pool-spreading model [\cref{eq:swe}].}
  \label{fig:model_overview}
\end{figure}

In the following, we present a model to predict the flow of LNG in a spill event following a containment breach.
The model accounts for important mechanisms such as boiling and evaporation, as well as enrichment of the LNG mixture due to evaporation, as shown in \cref{fig:model_overview}.

When the LNG is spilled on water due to LOC, it will spread and start to evaporate.
Vapor bubbles are continuously formed at the water--LNG interface and rise to the surface.
The bubbles displace the liquid phase, which reduces the average density of the liquid--vapor mixture.
The reduced density implies that the volume and thus the height of the LNG--air surface increases.
This in turn affects the spreading rate that depends on the leading-edge height~\citep{fyhn2019}.
Because of the coupling between spreading rate and vapor volume, the flow model must include the dynamics of the vapor mixed into the LNG.

Throughout this section, we consider axisymmetric spill geometries where the local fluid states are depth-averaged at each position throughout the pool.
Because of this, all properties are functions of time~$t$ and radius~$r$.
We take the origin $r=0$ to be the center of the spill source, and $t=0$ to be the start of the spill event.

We assume that the liquid mass is much larger than the mass of the vapor bubbles entrained in the LNG, and that the density of this vapor is nearly constant.
We further assume that the pressure is atmospheric, $p = p_\atm$, and thermal equilibrium such that the LNG temperature is always at the bubble point, that is, $T = T_\bub(p_\atm, \vect z)$, where $\vect z$ is the LNG component mass-fraction vector which changes with time.

\subsection{Pool spreading}
\label{sub:pool}
The dynamics of an LNG pool spreading on water may be captured accurately by the two-layer shallow-water equations (SWE)~\citep{audusse11,ungarish2013}.
These may be derived from the Euler equations for two liquid layers by assuming a negligible vertical velocity within each layer.
It is also common to use various effective forms of the single-layer SWE modified to represent the two-layer system, see \eg Refs.~\citep{verfondern2007,hatcher2014}.
In this work, we use a form of the single-layer SWE derived and analysed by \citet{fyhn2019}.
This form of the SWE has the benefit of being simple while still capturing essential features of the two-layer equations for liquid-on-liquid spreading.
In particular, it can be used without any additional boundary condition for the leading edge of the spill, as long as the water depth is large relative to the thickness of the LNG layer.

We employ the two-dimensional SWE to model the flow of an LNG liquid--vapor mixture, that is,
\begin{subequations}
  \label{eq:swe}
  \begin{align}
    &\pd t m_i + \div (m_i \vect u) = \dot{m}_i^\spill + \dot{m}_i^\evap, \\
    \label{eq:vap_cons}
    &\pd t h_\vap + \div (h_\vap \vect u)
      = \dot h_\vap^\evap + \dot h_\vap^\release, \\
    \label{eq:swe-momentum}
    &\pd t \vect u + (\vect u \cdot \nabla)\vect u +  g_\eff \nabla h = 0.
  \end{align}
\end{subequations}
Here $m_i$ is the liquid mass per area of component $i$,
$\vect u$ is the horizontal flow velocity,
$h_\vap$ is the vapor volume per area,
$\dot m_i^\spill$ is the spill source term,
$\dot m_i^\evap$ represents the liquid mass loss due to evaporation,
$\dot h_\vap^\evap$ represents the increase in the vapor phase due to evaporation,
$\dot h_\vap^\release$ represents the decrease in the vapor phase due to vapor bubbles released to atmosphere,
$h$ is the thickness of the liquid--vapor mixture,
and $g_\eff = \delta g$ is the effective gravitational acceleration.
$\delta$ is the buoyancy factor,
\begin{equation}
  \delta = \frac{\rho_\water - \rho}{\rho_\water},
  \label{eq:delta}
\end{equation}
where $\rho_\water$ is the water density and $\rho$ is the overall density of the liquid--vapor mixture,
\begin{align}
  \rho = \frac{\sum_i m_i}{h}.
  \label{eq:rho_mix}
\end{align}
Here we have assumed that the vapor mass is negligible, which is a good assumption as long as there is no foaming.
The liquid--vapor mixture thickness is given by
\begin{equation}
  h = \frac{\sum_i m_i}{\rho_\liq} + h_\vap,
\end{equation}
where $\rho_\liq = \rho_\liq (T_\bub, p_\atm, m_i)$ is the liquid LNG density.
Both $\rho_\liq$ and $T_\bub$ are calculated from the EoS, see \cref{sec:thermo}.

As shown by \citet{fyhn2019}, the water--LNG and LNG--air interface positions are given by
\begin{align}
  \label{eq:hwater}
  h_\water &= -(1-\delta)h, \\
  \label{eq:hLNG}
  h_\text{LNG} &= \delta h,
\end{align}
where the reference $h_\water=0$ is the water level where there is no LNG.

\subsection{Evaporation}
\label{sub:evap}
LNG spilled on water immediately begins to evaporate due to the large temperature difference.
Throughout this work, we assume that the absorbed heat from the water into the LNG goes to evaporation and to heating of the LNG as the boiling point is shifted.
Evaporation produces vapor that passes through the LNG before it is released.
The effect of evaporation is represented in the flow model~\eqref{eq:swe} through the source terms $\dot m_i^\evap$, $\dot h_\vap^\evap$, and $\dot h_\vap^\release$.

To estimate $\dot m_i^\evap$, we will assume that the composition of the evaporating gas is that of the \emph{incipient vapor phase} at the bubble point.
The change in mass of component $i$ in the liquid due to evaporation is given by
\begin{equation}
  \dot m_i^\evap = \pdtot {m_i} H \pdd H t,
\end{equation}
where $H$ is the specific enthalpy.
To calculate $\pdtoti {m_i} H$ numerically, we start from a saturated liquid state, add a small amount of enthalpy $\Delta H$, then calculate the corresponding liquid--vapor state from the EoS, see \cref{sec:thermo}.
Based on this state, we find the change of liquid mass for each component~$\Delta m_i$, which is always negative.
Next, since the pool is a system at constant pressure, addition of heat causes a corresponding increase of the enthalpy.
This implies that
\begin{equation}
  \pdd H t = \frac{\dot{q}}{\sum_i m_i},
\end{equation}
where $\dot q$ is the heat flux, which is further discussed in \cref{sec:heatflux}.

The vapor volume generated per area is given directly from the evaporation rate,
\begin{equation}\label{eq:vol_evap}
  \dot h_\vap^\evap = \frac{\sum_i \dot m_i^\evap}{\rho_\vap},
\end{equation}
where $\rho_\vap$ is the vapor density as predicted by the EoS.
The vapor bubbles rise through the pool due to buoyancy at an average velocity~$u_B$.
We have used $u_B=\SI{24}{cm/s}$, since this was measured by \citet{chang1983} for pure methane.
The timescale for bubble migration through the pool is relatively short, so we assume that the vapor bubbles are immediately dispersed uniformly throughout the liquid--vapor mixture.
In this case, the release rate of vapor volume per surface area is given by
\begin{equation}
  \dot h_\vap^\release = u_B \frac{h_\vap}{h}.
  \label{eq:vol_release}
\end{equation}

Due to the short timescale of bubble migration compared to the timescale of the spill flow, a quasi-steady state is quickly established where the rate of vapor released from the top of the pool is equal to the rate of vapor added at the bottom due to evaporation, \ie $\dot h_\vap^\evap = \dot h_\vap^\release$.
In this state, the vapor volume-fraction in the pool must be
\begin{equation}
  \frac{h_\vap^\text{max}} h = \frac{\dot h_\vap^\evap}{u_B}.
\end{equation}
We further obtain a minimal mixture density which is the same as that derived by \citet{chang1983},
\begin{equation}
  \label{eq:rhomin}
  \rho^\text{min} = \rho_l\left(1 - \frac{\dot h_\vap^\evap}{u_B}\right).
\end{equation}

When the LNG film is thin, the timescale of bubble migration is small compared to the timescale of the pool spreading.
We therefore assume $h_\vap = h_\vap^\mathrm{max}$ in the regions where $h < \SI{1}{\cm}$.

\subsection{Heat-transfer model}
\label{sec:heatflux}
The heat flux between a heat source and a boiling liquid strongly depends on the temperature difference.
For large temperature differences, which is typically the case for LNG boiling on water~\citep{boe1996}, the heat transfer is within the film-boiling regime.
When the methane fraction in the LNG decreases due to evaporation, the temperature difference drops gradually, and at some point the boiling enters the transition boiling regime.
However, as will be explained in \cref{sub:triggering}, a change of boiling regime from film boiling to transition boiling is thought to be part of the mechanism for RPT triggering.
We therefore restrict the heat-transfer model to the film-boiling regime, since the main goal is to investigate the onset of RPT triggering.

Standard film-boiling heat-transfer correlations, such as those by \citet{berenson1961} and \citet{kalinin1975}, tend to underpredict the heat transfer between light hydrocarbons and metallic heaters.
Furthermore, mixture effects on boiling are known to be significant, even for nearly pure fluids \citep{yue1973, valencia1979}.
Experimental results on the boiling regimes of LNG \citep{brown1968} and pure methane \citep{sciance1966phd} produce very different boiling curves.
When the heater material is changed from a metal plate to water, the heat-transfer coefficient increases by a factor of~1--4~\citep{boe1996, valencia1979, parnarouskis1980, drake1975b, sciance1966phd, brown1968}.
There are several factors that can contribute to this, such as increased LNG to heater area because of the uneven interface or big changes in the thermal transport properties of the heater.
Further experimental investigations are necessary to reduce this significant uncertainty.

In this work, we have used a film-boiling model developed by \citet{sciance1967film} for light hydrocarbons,
\begin{equation}
  \label{eq:sciance}
  \dot q = c_1\frac{k_\vapfilm}{l_\capillary}
    \left[\frac{\Ra\, \Delta H'}{T_r^2 c_{p,\vap}
    \Delta T}\right]^{c_2},
\end{equation}
with
\begin{subequations}
  \begin{align}
    \Delta H' &= \Delta H^\evap + c_{p,\vap}\Delta T/2, \\
    \Ra &= \frac{g \rho_\vapfilm \left(
        \rho_\liq - \rho_\vapfilm
    \right) l_\capillary^3 c_{p,\vapfilm}}{\mu_\vapfilm k_\vapfilm}, \\
    l_\capillary &= \sqrt{\frac{\sigma}{g\left(\rho_\liq -\rho_{\vap}\right)}}.
  \end{align}
\end{subequations}
Here $c_1$ and $c_2$ are model parameters, $k$ is the thermal conductivity, $l_\capillary$ is the capillary length, $\Ra$ is the two-phase Rayleigh number, $\Delta H'$ is the difference in specific enthalpy between the LNG and the vapor film, $T_r=T_\bub/T_c$ is the reduced temperature of the mixture, $T_c$ is the critical point temperature, $c_p$ is the isobaric specific heat, $\Delta T = T_\water - T_\bub$ is the LNG--water temperature difference, $\Delta H^\evap$ is the specific enthalpy of evaporation, $g$ is the acceleration of gravity, $\rho$ is the density, $\mu$ is the dynamic viscosity, and $\sigma$ is the interface tension of the LNG liquid--vapor interface.
The subscripts $\liq$ and $\vap$ indicate that a property should be evaluated for the LNG liquid and vapor phase at the bubble point, respectively.
The subscript $\vapfilm$ indicates that a property should be evaluated at the vapor-film temperature.
The model parameters used by \citet{sciance1967film} are $c_1=0.369$ and $c_2=0.267$.
We found that $c_2=0.3$ allowed us to better fit LNG--water heat-flux data \citep{boe1996, valencia1978, burgess1972, drake1975}.

The thermodynamic properties and the transport properties, \ie the viscosity, thermal conductivity, and surface tension, are described in further detail in \cref{sec:thermo}.

\subsection{Prediction of RPT triggering}
\label{sub:triggering}
\begin{figure}[btp]
  \centering
  \includegraphics[width=0.5\textwidth]{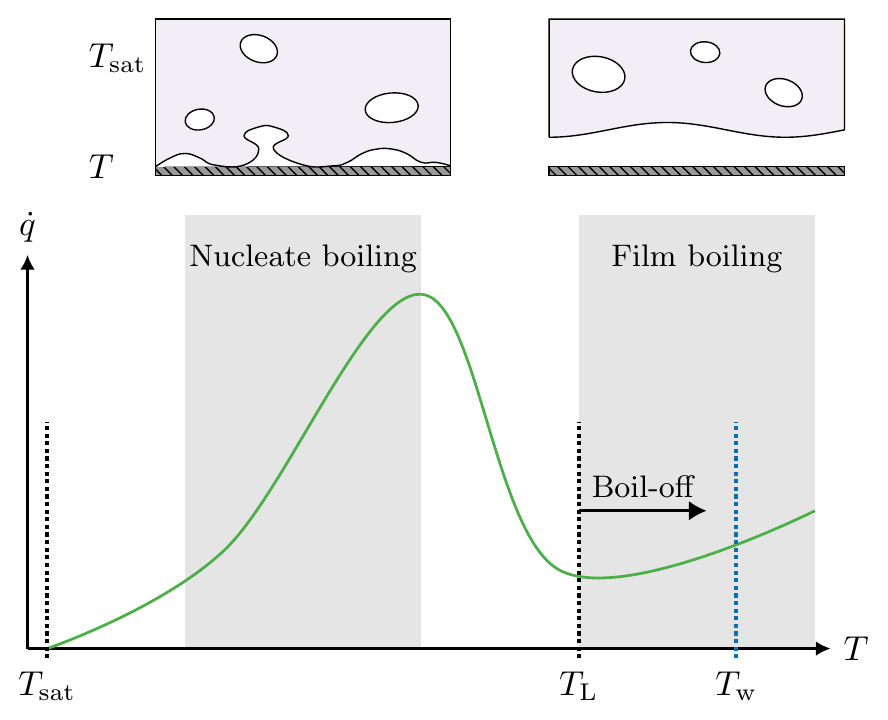}
  \caption{Illustration of the boiling curve for saturated LNG pool-boiling.
  This is a plot of boiling heat flux~$\dot{q}$ against a variable surface temperature~$T$.
  Of particular interest is the \emph{Leidenfrost temperature}~$T_\leid$, which marks the lower end of the film-boiling regime.
  Also shown is the typical temperature of seawater~$T_\water$, which is significantly above the Leidenfrost temperature of LNG. However, during boil-off (removal of methane) the Leidenfrost point will shift to the right, eventually crossing the water temperature (film-boiling collapse).}
  \label{fig:boiling_curve_RPT}
\end{figure}

The fundamental triggering condition for delayed RPT is dealt with in our previous work \citep{aursand2018b}.
It may be summarized by the following \emph{triggering window},
\begin{equation}
  T_\shl < T_\water < T_\leid,
  \label{eq:triggering_criterion}
\end{equation}
where $T_\shl$ is the superheat limit temperature of LNG, $T_\water$ is the temperature of seawater, and $T_\leid$ is the Leidenfrost temperature of LNG.
This criterion is deceptively simple, as it appears to only require the evaluation of three numbers.
However, both $T_\shl$ and $T_\leid$ depend on the LNG composition.
This criterion is \emph{not} satisfied for typical LNG compositions.
This is due to the fact that both $T_\leid$ and $T_\shl$ are far below the water temperature.
However, as boil-off proceeds, both $T_\leid$ and $T_\shl$ gradually increase while maintaining $T_\shl < T_\leid$.
This eventually causes \cref{eq:triggering_criterion} to be satisfied, in which case RPT may be triggered.
\Cref{fig:boiling_curve_RPT} shows the \emph{boiling curve} of LNG and indicates how boil-off shifts $T_\leid$ towards $T_\water$.
As $T_\shl < T_\leid$ is maintained during the boiling process, calculating $T_\shl$ is not required.

Note that actual triggering of RPT is not part of the spill model.
That is, when the RPT criterion is satisfied, the spill simulation is continued as if RPT did not occur.
Instead, we track the amount of LNG that satisfies the triggering criterion, that is,
\begin{equation}
  M_\RPT(t) = \int\sum_i m_i^\RPT(t,r) \; 2\pi r\dd r,
  \label{eq:M_rpt}
\end{equation}
where
\begin{equation}
  m_i^\RPT(t,r) = \begin{cases}
    m_i(t,r) & \text{if } T_\leid[\vect m(r,t)] >  T_\water, \\
    0        & \text{else,}
  \end{cases}
\end{equation}
where $\vect m = (m_1, \dots, m_n)$ is the mass vector of the $n$ LNG components.
The Leidenfrost temperature $T_\leid$ is defined in \cref{sec:thermo}.

\subsection{Thermodynamics and transport properties}
\label{sec:thermo}
An \eos is used to predict the thermodynamical properties of the mixture at a given state.
In this work, we use an extended corresponding states \eos~\citep{michelsen2007}, since it represents a good compromise between accuracy and computational speed~\citep{wilhelmsen2016-thermopack}.
We use the Peng--Robinson \eos~\citep{peng1976} to calculate the shape factors and methane as a reference fluid described with the modified Benedict Webb Rubin (MBWR) EoS~\citep{younglove1987}.
The accuracy of the EoS is good for mixtures with a high methane content.
For the predicited densities and heat-capacities of pure methane, the error is below \SI{2}{\percent} for gas and supercritical states, and approximately \SI{4}{\percent} in the liquid region when compared to GERG-2008~\citep{wilhelmsen2016-thermopack}.

For the selected \eos, multiple implicit thermodynamic properties must be determined numerically.
Especially, the LNG bubble temperature~$T_\bub$ is predicted at atmospheric pressure for a given composition of $n$ components $(m_1, \dots, m_n)$.
At specified atmospheric pressure, specific enthalpy, and total composition, the thermodynamically stable phase distribution and temperature must be calculated.
The latter follows from a $pHz$ flash, where the solution is found iteratively through an implicit relation, see Ref.~\citep{michelsen2007} for details.

The Leidenfrost temperature of a fluid depends not only on the fluid properties, but also thermal transport properties of the heater \citep{baumeister1973}, geometry and interface roughness \citep{bernardin1999}.
For pure fluids using the van der Waals \eos, \citet{spiegler1963} derived a simple relation that relates the Leidenfrost temperature at sub-critical pressures to the critical temperature: $T_\leid = (27/32) T_\crit$.
This approach was also adopted by \citet{aursand2018b}, who showed that this simple correlation gave an excellent prediction for methane, and generally a decent upper estimate for other hydrocarbons.
Due to lack of data, the same correlation was used for mixtures, without any validation.
As $(27/32) T_\crit$ corresponds to the liquid spinodal of the van der Waals \eos when $p \rightarrow 0$, we have chosen to correlate $T_\leid = T_\spinodal(\vect m, p)$ in this work.
The method produces a similar composition dependence as that reported by \citet{yue1973} for binary mixtures.
The liquid spinodal was calculated as the temperature where the smallest eigenvalue of the Gibbs energy Hessian matrix with respect to mole numbers become zero \citep{aursand2017,gjennestad2020}.

Since the liquid is assumed to always be saturated, the thermodynamical properties of the vapor phase are calculated from the EoS by finding the incipient vapor phase at the corresponding temperature and pressure.

The viscosity and thermal conductivity are calculated using the TRAPP method by \citet{ely1981, ely1983}, with propane as a reference fluid.
The surface tension is calculated using the corresponding-state correlation presented by \citet{poling2001}.
These models take density as an input, and so the EoS influences the accuracy of the transport properties.
Typically, we get errors less than \SI{5}{\percent} for the transport properties for both the liquid and vapor phase.

\subsection{Numerical implementation}
In the axisymmetric case, the SWE \eqref{eq:swe} may be reduced to
\begin{subequations}
  \label{eq:swe_axisymmetric}
  \begin{align}
    \pd t m_i + \pd r (m_i u)
      &= \dot m_i^\spill + \dot m_i^\evap - \frac{m_i u}{r}, \\
    \pd t h_\vap + \pd r(u h_\vap)
      &= \dot h_\vap^\evap + \dot h_\vap^\release - \frac{u h_\vap}{r}, \\
    \pd t u + \pd r \left( \frac{1}{2} u^2 + g_\eff h \right) &= 0.
  \end{align}
\end{subequations}
These equations are discretized spatially using a finite-volume scheme.
We employ the FORCE (first-order centered) flux~\citep{toro2000} and the second-order MUSCL (Monotonic Upstream-Centered Scheme for Conservation Laws) reconstruction with a minmod limiter~\citep{leveque02} in each finite volume.
The solutions are advanced in time with a standard third-order three-stage strong stability-preserving Runge--Kutta method~\citep{ketcheson05}.

The time steps are restricted by the Courant–Friedrichs–Lewy (CFL) condition with a CFL number of \num{1.0} for all cases such that
\begin{equation}
  \Delta t \le \frac{\Delta x}{\max_j |u_j| + \max_j \sqrt{g_\eff h_j}},
\end{equation}
where subscript $j$ indicates grid cell $j$.
In addition, we also restrict the time steps to account for the spill source term,
\begin{equation}
  \Delta t \le \frac 1 2 \frac{\max_\text{cells in spill region} \sum_i m_{i,j}}{\sum_i \dot m_{i,j}^\spill},
\end{equation}
The denominator is regularized to avoid division by zero in \eg scenarios with no initial LNG.

The numerical simulations are run on uniform axisymmetric 1D meshes with \SI{100}{grid\ points\per\meter}.
This was found to be sufficient to ensure grid independence for the simulation results.

Unless otherwise specified, we apply zero-gradient boundary conditions for all variables at the boundaries, except that we that we specify $u = 0$ at $r = 0$.

\section{Analytical estimates}
\label{sec:analytical}
Detailed models such as the one presented in \cref{sec:spill_model} are useful for improving our understanding of fundamental phenomena.
However, for practical applications such as safety assurance, a simplified model is desired.
Such models are easier to understand and apply, require fewer input parameters, and often produce satisfactory results.
In this section, we develop a method to predict the radius of RPT for steady-state spills of LNG.
The method is used to derive an analytical estimate of the RPT radius and triggering time for an axisymmetric two-dimensional tank spill of LNG on water.
This spill scenario could be the result of an operational failure in a loading arm during loading or off-loading of an LNG tanker to an on-shore facility.

We define the \emph{boil-off limit} as,
\begin{equation}
  \theta \equiv 1 - \sum_i \tilde{z}_i,
  \label{eq:boil-off-limit}
\end{equation}
where $\tilde{z}_i$ is the mass fraction of chemical component~$i$ at RPT triggering measured relative to the \emph{initially} spilled mass.
Thus, $\theta$ is the mass fraction of the initially spilled LNG that must evaporate before an RPT triggering can occur.
This boil-off limit only depends on the spill composition and evaporation model, and can be determined in many different ways.
Here we calculate it numerically from the condition $T_\leid(\tilde{\vect z}) = T_\water$, which ensures that the RPT triggering criterion from \cref{sub:triggering} is satisfied.
Alternatively, it can be analytically estimated to within a few percent from the spill composition alone using the correlations derived by \citet{aursand2018b}.
It is also possible to measure it experimentally by placing LNG into a container with enough water, and record how its weight changes until triggering.
It is worth noting that if only methane evaporates from the LNG mixture, the boil-off limit can also be written in the simpler form $\theta = z_1 - \tilde{z}_1$, where $z_1$ and $\tilde{z}_1$ are the methane mass fractions at spill and triggering, respectively.

Our model from \cref{sec:spill_model} and the following analysis is applicable to any multicomponent mixture.
However, for simplicity, we will henceforth consider LNG mixtures~$\vect z = (z_1, z_2, z_3)$ that have three components: methane~(\ce{CH4}), ethane~(\ce{C2H6}), and propane~(\ce{C3H8}).
We consider three such mixtures, which we denote $\vect z_\text{A}$, $\vect z_\text{B}$, and~$\vect z_\text{C}$.
The assumed compositions and calculated boil-off criteria for these mixtures are listed in \cref{tab:compositions}.
Thus, similar to previous work in the topic~\citet{horvat2018}, we have omitted nitrogen from the compositions.
With a small amount of nitrogen ($\le\SI{0.5}{\percent}$), we have used the complete model to confirm that this has a negligable influence on the results.

\begin{table}[b!]
  \centering
  \caption{List of initial LNG compositions and the calculated boil-off limit for RPT~$\theta$.}
  \label{tab:compositions}
  \begin{tabular}{ccccc}
    \toprule
    Composition & \ce{CH4} (\si{\masspercent}) & \ce{C2H6} (\si{\masspercent})
                & \ce{C3H8} (\si{\masspercent}) & $\theta$ (\si{\percent}) \\
    \midrule
    $\vect z_\text{A}$ & 90.0 & \phantom{0}7.5 & 2.5 & 89.1 \\
    $\vect z_\text{B}$ & 80.0 & 15.0           & 5.0 & 78.1 \\
    $\vect z_\text{C}$ & 70.0 & 22.5           & 7.5 & 67.2 \\
    \bottomrule
  \end{tabular}
\end{table}

\begin{figure}[tbp]
  \centering
  \tikzsetnextfilename{theta-boiloff}
  \begin{tikzpicture}[font=\small]
    \begin{axis}[
      height=0.4\textwidth,
      width=0.5\textwidth,
      xlabel = {Boil-off (\si{\percent})},
      ylabel = {$T_\text{L}$ (\si{\celsius})},
      xmin = 0,
      xmax = 100,
      ymin = -100.0,
      ymax = 25.0,
      x axis line style=-,
      legend pos = north west,
      table/y = T,
      ]
      \addplot[cl3-1] table {data/theta-boiloff-70.0.csv};
      \addplot[cl3-2] table {data/theta-boiloff-80.0.csv};
      \addplot[cl3-3] table {data/theta-boiloff-90.0.csv};
      \addlegendentry{$\vect z_\text{C}$ (\SI{70}{\masspercent} methane)}
      \addlegendentry{$\vect z_\text{B}$ (\SI{80}{\masspercent} methane)}
      \addlegendentry{$\vect z_\text{A}$ (\SI{90}{\masspercent} methane)}

      \addplot[gray, thin, densely dashed] coordinates {(0,0) (100, 0)};

      \coordinate (p1) at (axis cs: 67.18, 0);
      \coordinate (p2) at (axis cs: 78.12, 0);
      \coordinate (p3) at (axis cs: 89.06, 0);
      \coordinate (tw) at (axis cs: 100, 0);
    \end{axis}

    \fill (p1) circle(2pt);
    \fill (p2) circle(2pt);
    \fill (p3) circle(2pt);

    \draw[thin, densely dotted] (p1) -- +(-0.3, 1.5) node[above] {$\theta = \SI{67.2}{\percent}$};
    \draw[thin, densely dotted] (p2) -- +( 0.6, 1.0) node[above] {$\theta = \SI{78.1}{\percent}$};
    \draw[thin, densely dotted] (p3) -- +( 0.9, 0.5) node[above, xshift=0.3cm]
      {$\theta = \SI{89.1}{\percent}$};

    \node[right] at (tw) {$T_\water$};
  \end{tikzpicture}
  \caption{The Leidenfrost temperature as a function of boil-off for three different initial mixtures of methane, ethane, and propane.
    The initial ratio between ethane and propane is 3:1 for each curve.
    The intersection between the Leidenfrost temperature and the water temperature yields the boil-off limit~$\theta$ for an RPT event.
  Here we used a simplified boiling model where only the lightest component evaporates at any time.}
  \label{fig:t_leid-boiloff}
\end{figure}
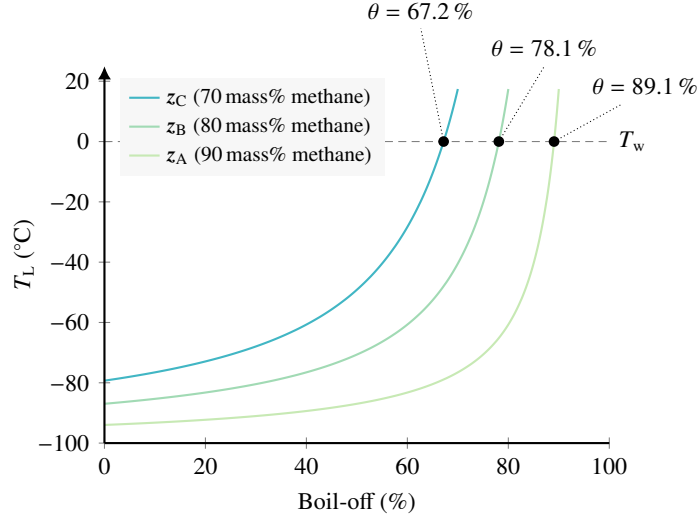

\subsection{Methodology for triggering prediction}
\begin{figure}[tb!]
  \centering
  \begin{subfigure}[t]{0.48\columnwidth}
    \centering
    \includegraphics[width=1.0\textwidth]{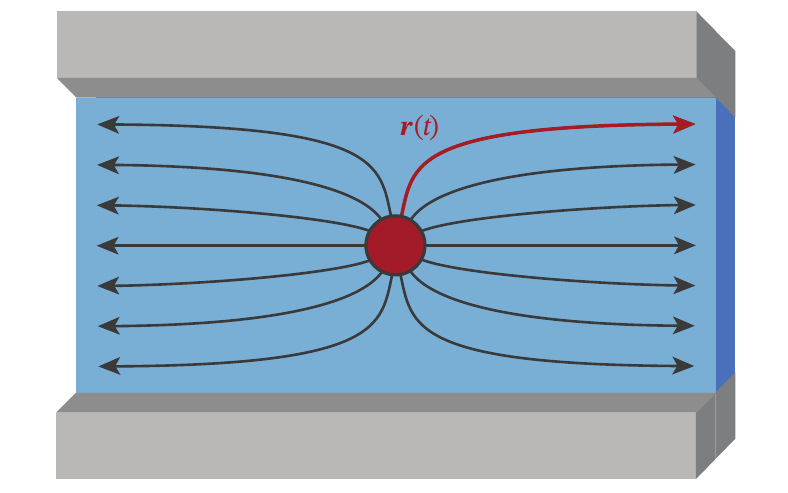}
    \caption{Paths $\vect r(t)$ taken by control masses at the spill source.}
    \label{fig:sketch-boiloff-a}
  \end{subfigure}
  \begin{subfigure}[t]{0.48\columnwidth}
    \centering
    \includegraphics[width=1.0\textwidth]{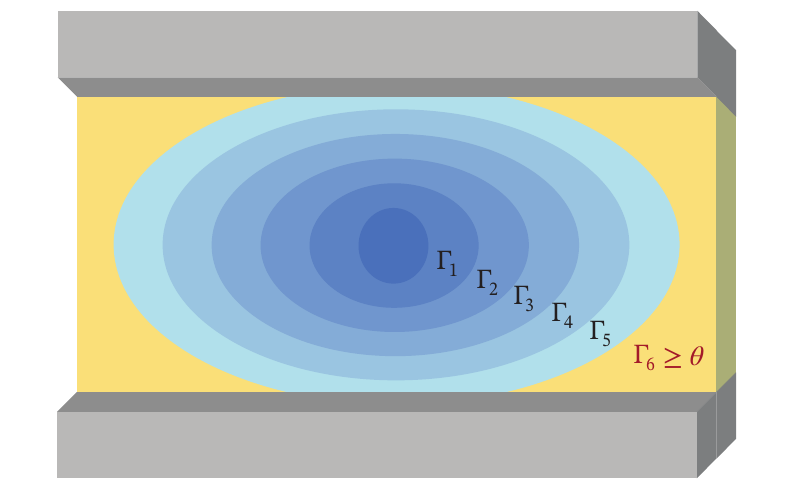}
    \caption{Contours for the corresponding boil-off fractions~$\Gamma(t)$.}
    \label{fig:sketch-boiloff-b}
  \end{subfigure}
  \caption{Sketch of the RPT prediction strategy for continuous spills.
    (a)~Here, we illustrate a cylindrical spill source in a wide channel, resulting in 2D spreading followed by 1D spreading.
    If the steady-state velocity field~$\vect u(r)$ is known, the path~$\vect r(t)$ that a control mass takes (red line) can be calculated by integrating $\vect u(\vect r)$ from a point $\vect r(0)$ on the spill boundary (red circle).
    By repeating this for each such point~$\vect r(0)$, all paths~$\vect r(t)$ from the source can be mapped out (gray lines).
    We can then also describe~$t(\vect r)$, \ie the time LNG spends moving from the source to any given point~$\vect r$.
    (b)~Since a roughly constant amount of methane evaporates per unit time the LNG is on sea water, the fraction~$\Gamma$ that has evaporated at each point~$\vect r$ can be calculated.
  By comparing the resulting contours~$\Gamma_n$ to the \emph{boil-off limit}~$\theta$, we can predict the region at risk of an RPT event~(yellow).}
  \label{fig:sketch-boiloff}
\end{figure}

Consider a spill scenario, such as a containment breach in a large tank or a rupture in an LNG transfer line, which causes LNG to leak onto sea water.
If the spill continues at a steady rate~$S$, the flow should approach an approximately steady-state pattern, as shown in \cref{fig:sketch-boiloff}.
In this limit, the LNG mass distribution~$m$ and velocity field~$\vect{u}$ become nearly time-independent.

Assume that the flow patterns are not affected by evaporation, that is, that the flow velocity $\vect u$ is the same regardless of evaporation.
Then we may replace $m$ in a case without evaporation with $(1-\Gamma)m$ in a corresponding case with evaporation, where $\Gamma$ is the fraction that has boiled off at each position.
This is usually a reasonable assumption since the \emph{amount of methane} that evaporates per unit area and time is roughly constant, so the \emph{methane fraction} changes faster where the LNG is spread thin, \ie far from the spill source.
This is also observed by numerically solving the more advanced model, as the chemical composition remains nearly constant throughout \emph{most} of the spill region (see \cref{fig:contspill_single_Tleid_a}).
Thus, if we neglect evaporation, the steady-state $m(\vect{r})$ and $\vect{u}(\vect{r})$ can be obtained \eg via computational fluid dynamics or analytical approximations based on mass conservation.

We now consider an \emph{arbitrary} spill geometry, and derive a general equation to predict when and where RPT events may occur.
This is achieved by decoupling the model of how LNG \emph{spreads} from the model of how it \emph{evaporates and eventually undergoes RPT}.
This is an approximation that can enable simple RPT predictions in realistic geometries.
The predictions are validated numerically in \cref{sec:contspill}.

Methane typically evaporates much faster than the other components in LNG.
We therefore approximate the specific enthalpy of evaporation~$\Delta H$ by the methane value~$\Delta H_1$, which is listed in standard chemical tables.
Furthermore, we assume that the heat flux~$\dot q$ from sea water into LNG is roughly independent of position and composition.
With these simplifications, the equations in \cref{sub:evap} yield the following methane evaporation rate
\begin{equation}
  \dot{m}_1^\evap \approx -\frac{ \dot{q} }{ \Delta H_1 }.
  \label{eq:contspill-m1evap}
\end{equation}
The evaporation rates of all other chemical species in the spilled LNG $\dot{m}_i^\evap$ have been neglected.
The enthalpy contribution of mixing methane with the LNG, is also neglected.

Given the above assumptions, we next track how a small \emph{control mass} flows through the system.
If the control mass starts at a point~$\vect{r}(0)$ at time~$0$, the path~$\vect{r}(t)$ it traverses is given by the steady-state velocity field~$\vect{u}(\vect{r})$:
\begin{equation}
  \vect{r}(t) - \vect{r}(0) = \int_0^t \frac{ \dd\vect{r} }{ \dd t } \dd t = \int_0^t \vect{u}[\vect{r}(t)] \,\dd t.
\end{equation}
This equation may be solved implicitly to find a specific position $\vect r(t)$ at any given time $t$.
It may be challenging to solve an implicit equation analytically, but it should be straightforward to solve numerically for any geometry.

As mentioned previously, the methane evaporation rate~$\dot m_1^\evap$ is assumed to be constant.
However, how fast this changes the \emph{composition} of LNG depends on position, since the mass per unit area~$m(\vect{r})$ is not constant.
We can relate the mass fractions of each component $z_i$ to the mass components as $m_i = m z_i$, which for methane implies that the change in composition is $\dot z_1 = \dot m_1^\evap / m$.
Note that we defined $\vect z$ using the mass profile~$m(\vect r)$ calculated \emph{in the absence} of evaporation.
Thus, as mass evaporates, this definition of $\vect z$ implies that the sum of components $\sum_i z_i < 1$, and the ``missing mass'' is the boil-off fraction~$\Gamma$ that plays a central role in RPT prediction below.

The total amount of methane that has evaporated from the control mass can then be found via integration of $\dot z_1 = \dot{m}_1^\evap / m$ along the path~$\vect{r}(t)$.
Moreover, we can substitute in \cref{eq:contspill-m1evap} for $\dot{m}_1^\evap$.
This leaves us with:
\begin{equation}
  \Delta z_1(t) = -\frac{\dot{q}}{\Delta H_1} \int_0^t \frac{\dd t}{m[\vect{r}(t)]}.
\end{equation}
This is negative since methane disappears during evaporation.
As \emph{only} methane evaporates, the \emph{boil-off fraction} is:
\begin{equation}
  \Gamma(t) = -\Delta z_1(t) = \frac{ \dot{q} }{ \Delta H_1 } \int_0^t \frac{\dd t}{m[\vect{r}(t)]}.
\end{equation}
As introduced in \cref{eq:boil-off-limit}, the time~$\tilde t_\RPT$ the control mass has to travel before an RPT event occurs is defined according to the \emph{boil-off limit} $\Gamma(\tilde t_\RPT) \equiv \theta$, which implicitly defines $\tilde t_\RPT$ for a given path~$\vect{r}(t)$.
The corresponding boundary to the region where an RPT event might occur is then given by $\vect{r}_\RPT \equiv \vect{r}(\tilde t_\RPT)$.
In both cases, the path~$\vect{r}(t)$ is defined by the initial position along the spill boundary~$\vect{r}(0)$;
to obtain the full RPT hazard region, one therefore has to repeat this consideration for each point~$\vect{r}(0)$ along the spill boundary.

To summarize, the region at risk of RPT can in general be determined from a known mass distribution~$m(\vect{r})$ and velocity field~$\vect{u}(\vect{r})$ by solving these equations for each point $\vect{r}(0)$ along the spill boundary:
\begin{align}
  \label{eq:r-follow}
  \vect{r}(t) &= \vect{r}(0) + \int_0^t \vect{u}[\vect{r}(t)] \,\dd t, \\
  \label{eq:boil-off-equation}
  \theta &= \frac{ \dot{q} }{ \Delta H_1 } \int_0^{\tilde t_\RPT} \frac{\dd t}{m[\vect{r}(t)]}.
\end{align}
These equations may be solved for the path~$\vect r(t)$ and the triggering time~$\tilde t_\RPT$.
The locations at risk of delayed RPT events are then given by~$\vect r_\RPT \equiv \vect r(\tilde t_\RPT)$.
Note that~$\tilde t_\RPT$ is not the time it takes from the initial spill until a delayed RPT is possible.
This time is typically longer, since the initial spreading will be restricted by the water.

In order to estimate the time~$t_\RPT$ from the initial spill starts until a delayed RPT is possible, we use the method of characteristics.
In the absence of evaporation, \cref{eq:swe} can be written
\begin{subequations}
  \begin{align}
    \pd t u + u\left(\hat{\vect{u}} \cdot \nabla\right) u
    + c \left(\hat{\vect{u}} \cdot \nabla\right) 2c &= 0, \\
    \pd{t}2c + c \left(\hat{\vect{u}} \cdot \nabla\right) u
    + u\left(\hat{\vect{u}} \cdot \nabla\right) 2c &=
    - cu\, (\div\hat{\vect u}),
  \end{align}
\end{subequations}
where $c = \sqrt{g_\eff h}$, $\hat{\vect u}$ is a unit vector along $\vect u$, and $u = \abs{\vect u}$ is the magnitude of $\vect u$.
Combining these equations, we get
\begin{equation}
  \pdtot{}{t}\left(u + 2c\right) = -cu\, (\div\hat{\vect u})
  \quad \text{along} \quad
  \pdtot{\vect r}{t} = (c+u)\,\hat{\vect u}.
  \label{eq:characteristics}
\end{equation}
From this we can connect the velocity at the leading edge to the steady-state velocity distribution.
The characteristics described by \cref{eq:characteristics} will move faster than the shock at the leading edge.
Hence, we can find an equation that relates the height and velocity at the shock with the height and velocity further back where the solution can be assumed to be better approximated by the steady state solution.
Integrating \cref{eq:characteristics} from $t_a$ to $t_b$, where $\vect u[\vect r(t_b)] = \vect u_\LE$, gives
\begin{equation}
  \left(1+\sqrt 2\right) u_\LE =
  u[\vect r(t_a)] + 2\sqrt{g_\eff h[\vect r(t_a)]} -
  \int_{t_a}^{t_b}cu \,(\div\hat{\vect u}) \dd{t}.
\end{equation}
Here we assumed that $u_\LE = \hat{\vect n} \cdot \vect u_\LE$, which from the Rankine--Hugoniot condition implies that $u_\LE = \sqrt{2g_\eff h_\LE}$, as shown in Ref.~\citep{fyhn2019}.
If we assume that $\vect u[\vect r(t_a)]$ and $h[\vect r(t_a)]$ are well approximated by the steady state solutions and $t_b-t_a$ is sufficiently small so that we may neglect the contribution from the integral, we can write the time from initial spill to a possible RPT as
\begin{equation}
  t_\RPT = \left( 1+\sqrt 2 \,\right) \int\limits_{\vect r(t)} \frac{\dd r}{u + 2\sqrt{g_\eff h}},
  \label{eq:t_rpt}
\end{equation}
where the integration contour $\vect r(t)$ starts at $\vect r(0)$ and ends at $\vect r_\RPT$.
In general, each point $\vect r(0)$ along the spill boundary can give a different value of $t_\RPT$, so \cref{eq:t_rpt} should be minimized with respect to $\vect r(0)$.

This result can be used to perform RPT predictions for any geometry, and can in practice be coupled to a computational fluid-dynamics simulation as a post-processing step.
The idea is also summarized in \cref{fig:sketch-boiloff}.
However, we note that a central assumption behind the results above was that evaporation is not very important for the flow pattern.
In geometries where evaporation effects are more dominant to the flow patterns, the above formulation may yield less accurate results.
This is further discussed in \cref{sec:contspill}.

\subsection{Application to an axisymmetric 2D spill}
To illustrate the methodology derived in the previous subsection, we will now apply it to an axisymmetric two-dimensional spill of LNG on the ocean.
Such a scenario could be the result of a containment breach in a long transfer line between an LNG carrier and a bunkering facility.

\begin{figure}[b!]
  \centering
  \includegraphics[width=0.6\textwidth]{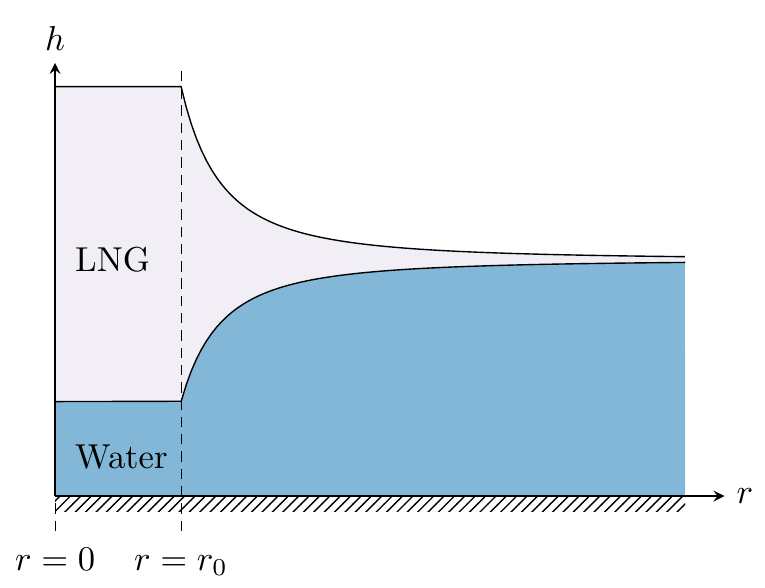}
  \caption{Sketch of an axisymmetric spill in steady state.}
  \label{fig:sketch-spill}
\end{figure}

The spill is assumed to occur within some radius~$r<r_0$ from the origin, see \cref{fig:sketch-spill}, while the earliest point an RPT event may occur is at some much larger radius~$r = r_\RPT$.
In the following, we first apply the above method to determine $r_\RPT$.
We then expand the analysis and derive estimates for the time~$t_\RPT$ when delayed RPT events first become possible.

In the absence of evaporation, mass conservation implies that the following relationship between the net spill rate~$S$, the radial velocity~$u$, and the mass~$m$ for $r > r_0$:
\begin{equation}
  \label{eq:mass-cons}
  2\pi r \, u(r) \, m(r) = S.
\end{equation}
This equation is solved for~$1/m$, which substituted into the boil-off equation~\eqref{eq:boil-off-equation} yields
\begin{equation}
  \frac{\dot q}{S \Delta H_1} \int_0^{t_\RPT} 2\pi r(t) \, u[r(t)] \, \dd t = \theta.
\end{equation}
However, $u \, \dd t = \dd r$ by definition, and $\int 2\pi r \, \dd r = \pi (r^2-r_0^2)$.
This can be used to rewrite the equation in terms of the position~$r_\RPT$.
Moreover, since we are interested in regions $r_\RPT \gg r_0$, we can also neglect the $r_0$ contribution.
This leaves us with an equation for the RPT location~$r_\RPT$:
\begin{equation}
  \pi r_\RPT^2 = S\theta \, \Delta H_1 / \dot{q} .
  \label{eq:contspill-r_rpt}
\end{equation}

This result requires only three input parameters to estimate $r_\RPT$: the heat flux~$\dot{q}$, spill rate~$S$, and boil-off limit~$\theta$.
It also has a straightforward physical interpretation:
It can be reformulated as $A_\RPT = \theta S/\mu$, where $\mu =  \dot{q}/\Delta H_1$ is the rate at which mass disappears per unit area.
The result $A_\RPT = \pi r_\RPT^2$ is simply the surface area required for a fraction $\theta$ of the spilled material~$S$ to evaporate, in accordance with the boil-off limit.
We can then easily solve for the boundary~$r_\RPT$ of the region~$A_\RPT$.
Beyond that boundary, the boil-off criterion is satisfied.

To estimate $t_\RPT$, we will find the steady-state velocity distribution $u(r)$.
By integrating the shallow-water equations~\eqref{eq:swe_axisymmetric}, it can be found that the steady-state height and velocity distributions must satisfy
\begin{subequations}
  \begin{align}
    \frac 1 2 u^2 + g_\eff h &= \varepsilon,
    \label{eq:steadystateVelocity} \\
    2\pi r \, \rho h u &= \int_0^r 2\pi r \, \dot m^\spill \dd t = \dot M^\spill,
    \label{eq:steadystateMass}
  \end{align}
\end{subequations}
where $\varepsilon$ is a constant to be determined, $\dot m^\spill$ is the spill rate density, and $\dot M^\spill(r) = S$ for $r \ge r_0$.
We solve \cref{eq:steadystateVelocity,eq:steadystateMass} to obtain a polynomial equation of degree three in $h$,
\begin{align}
  g_\eff h^3 - \varepsilon h^2 + \frac 12 \left(\frac{\dot M^\spill}{2\pi r\rho}\right)^2 = 0.
\end{align}
The solutions of a third-order polynomial equation can be found by Cardano's formula~\citep{abramowitz1972}.
From the three roots, we pick the desired solution to fit the boundary conditions, that the height at $r \to \infty$ should be smaller than that at $r = 0$ and that $u = 0$ at $r=0$.
We find that
\begin{align}
  h = \begin{cases}
    \displaystyle
    \frac{\varepsilon}{3g}\left[1 - 2\cos\left(\frac{\chi}{3} + \frac{2\pi}{3}\right)\right] &\text{ if } r < r_0, \\[10pt]
    \displaystyle
    \frac{\varepsilon}{3g}\left[1 - 2\cos\left(\frac{\chi}{3} - \frac{2\pi}{3}\right)\right] &\text{ if } r > r_0,
  \end{cases}
  \label{eq:hSteady}
\end{align}
where
\begin{align}
  \chi = \arccos\left[\frac{27\left(g_\eff\dot M^\spill\right)^2}{4\varepsilon^3 (2\pi r\rho)^2} - 1\right].
\end{align}
Next, we can determine $\varepsilon$ by demanding that the solution is continuous at $r = r_0$.
This is the case if $\chi(r_0) = 0$, which happens when
\begin{equation}
  \sqrt{2\varepsilon} = \sqrt[3]{\frac{\sqrt{27}S g_\eff}{2\pi r_0 \rho}}.
\end{equation}

We are now in a position to estimate $t_\RPT$ from \cref{eq:t_rpt}.
If $r_\RPT \gg r_0$, the integral will be dominated by its large-$r$ contributions.
Taylor expanding the integrand to first order in $r_0/r$ gives
\begin{equation}
  \frac{1}{u + 2\sqrt{g_\eff h}} =
  \frac{1}{\sqrt{2\varepsilon}} \left\{1-\frac{2}{3^{3/4}} \sqrt{\frac{r_0}{r}} +
  \frac{5}{3^{3/2}} \left( \frac{r_0}{r} \right) +
  \mathcal O\left[\left(\frac{r_0}{r}\right)^{3/2}\right]\right\}.
\end{equation}
If we now integrate this from $r_0$ to $r_\RPT$, we get
\begin{equation}
  \label{eq:contspill-t_rpt}
  t_\RPT \approx \frac{1+\sqrt{2}}{\sqrt{2\varepsilon}}\, f(r_\RPT/r_0)\, r_\RPT ,
\end{equation}
where we have written the result in terms of a function
\begin{equation}
  f(R) \equiv 1-\frac{4}{3^{3/4}}\frac{1}{\!\!\sqrt{R}} + \frac{5}{3^{3/2}}\frac{\ln R}{R},
\end{equation}
with $R = r_\RPT/r_0$.
Note that for $R \rightarrow \infty$, $f(R) \rightarrow 1$.
In this limit, $t_\RPT \sim r_\RPT / u_\infty$, where $u_\infty = \sqrt{2\epsilon}$ is the steady-state velocity for $R \rightarrow \infty$, giving an intuitive ``distance over speed'' result for~$t_\RPT$.
For finite~$R$, $f(R)$ includes more low-$R$ corrections.

The analytical solution we derived for~$h(r)$ can also be used to estimate the mass distribution~$m(r)$.
For $r>r_\RPT$, a fraction~$\theta$ of the spilled LNG has already evaporated, so the mass distribution is $m(r) \approx (1-\theta) \rho h(r)$, where $h(r)$ is the solution derived in the absence of evaporation and is given by \cref{eq:hSteady}.
By integrating the resulting mass per radial shell $2\pi r \, m(r)$ from the risk boundary~$r_\RPT$ to the leading edge of the spill~$r_\LE$, one can also estimate the total mass~$M_\RPT$ at risk of RPT.
This may then be combined with the analytical results for the explosive pressure and energy derived by \citet{aursand2018b} to obtain a worst-case scenario analysis.

A similar analysis is also applicable to one-dimensional spills, which approximate spills into narrow channels.
The mass conservation equation then changes to $2d\,u(x)\,m(x) = S$, where $d$ is the channel width and $x$ the coordinate along the channel, but the derivations follow along the same lines.

\section{Numerical results and discussions}
\label{sec:contspill}
We still consider an axisymmetric two-dimensional spill similar to what was discussed above.
This spill scenario will be used to demonstrate the applicability of the LNG spill model~\eqref{eq:swe_axisymmetric} and to compare the analytical estimates of $r_\RPT$~[\cref{eq:contspill-r_rpt}] and $t_\RPT$~[\cref{eq:contspill-t_rpt}] with simulation results from the full coupled model.

We assume that an LNG volume $V_0=\SI{10}{\meter\cubed}$ with composition $\vect z_\text{A}$ (\cref{tab:compositions}) is released onto sea water at a constant rate~$S$ through a hole with radius $r_0=\SI{0.1}{\meter}$.
The spill lasts until $t_s = \SI{30}{\second}$.
The initial density of the LNG at atmospheric pressure and bubble temperature is $\rho_0 = \SI{437}{\kg /\meter \cubed}$.
The spill rate is therefore $S = \SI{146}{\kg /\second}$.
The spill source term is set to
\begin{equation}
  \dot m_i^{\text{spill}}(r,t) = \begin{cases}
    {S_i}/{\pi r_0^2} & \text{when $r<r_0$ and $t<t_s$,} \\
    0 & \text{otherwise},
  \end{cases}
\end{equation}
where $S_i$ is the spill rate of component $i$.
Otherwise, the initial conditions are all zero.
The simulations are run in a domain of radius $\SI{60}{\meter}$.

Although this scenario is theoretical, it may still represent a semi-realistic accident in which LOC occurs on a loading arm during ship-to-ship bunkering of an LNG-fuelled ship.
An emergency shut-down valve may be expected to trigger after some time $t_s$ to restrict the total spill amount.
The values for the spill volume $V_0$ and duration $t_s$ are based on the case descriptions from \citet{ten-t}.

% Delta H/qdot = 5.1e5/6.9e4 = 7.391311
\pgfmathsetmacro{\predictedRadius}{r_rpt(0.891, 146, 7.391311)}
\pgfmathsetmacro{\predictedTime}{t_rpt(0.891, 146, 7.391311, 437.0, 0.1)}
The analytical estimates for $r_\RPT$ and $t_\RPT$ from \cref{eq:contspill-r_rpt,eq:contspill-t_rpt} require values for $\theta$, $\dot q$, and $\Delta H$.
The spilled LNG composition corresponds to $\theta = \SI{89.1}{\percent}$, as listed in \cref{tab:compositions}.
The heat flux $\dot q$ can be estimated from \cref{eq:sciance}, we find $\dot q \approx \SI{6.9e4}{\watt\per\meter\squared}$.
The specific enthalpy of evaporation can be found for methane from standard chemical tables, here we use $\Delta H^\evap \approx \SI{510}{\kilo\joule\per\kilogram}$.
With this, we calculate the estimates $r_\RPT \simeq \SI[round-mode = places, round-precision = 1]{\predictedRadius}{\meter}$ and $t_\RPT \simeq \SI[round-mode = places, round-precision = 1]{\predictedTime}{\second}$.

\begin{figure}[tbp]
  \centering
  \begin{subfigure}[t]{0.45\columnwidth}
    \centering
    \tikzsetnextfilename{contspill-single-h1}
    \begin{tikzpicture}
    \begin{axis}[
        xlabel={$r$ [\si{\meter}]},
        ylabel={$h$ [\si{\centi\meter}]},
        xmin=0,
        xmax=21,
        ymin=-8,
        ymax=32,
        table/y = h_bottom,
      ]
      \addplot[cl4-4]       table {data/contspill_t01.csv};
      \addplot[cl4-3]       table {data/contspill_t05.csv};
      \addplot[cl4-2]       table {data/contspill_t10.csv};
      \addplot[cl4-1, emph] table {data/contspill_t30.csv};

      \addlegendentry{$t=\phantom{1}\SI{1}{\second}$}
      \addlegendentry{$t=\phantom{1}\SI{5}{\second}$}
      \addlegendentry{$t=\SI{10}{\second}$}
      \addlegendentry{$t=\SI{30}{\second}$}

      \pgfplotsset{table/y = h_top}
      \addplot[cl4-4]       table {data/contspill_t01.csv};
      \addplot[cl4-3]       table {data/contspill_t05.csv};
      \addplot[cl4-2]       table {data/contspill_t10.csv};
      \addplot[cl4-1, emph] table {data/contspill_t30.csv};
    \end{axis}
    \end{tikzpicture}
    \caption{While the spill is ongoing.}
    \label{fig:contspill_single_h_a}
  \end{subfigure}
  \begin{subfigure}[t]{0.45\columnwidth}
    \centering
    \tikzsetnextfilename{contspill-single-h2}
    \begin{tikzpicture}
    \begin{axis}[
        xlabel={$r$ [\si{\meter}]},
        ylabel={$h$ [\si{\centi\meter}]},
        xmin=0,
        xmax=21,
        ymin=-0.4,
        ymax=0.64,
        table/y = h_bottom,
      ]
      \addplot[cl4-1, emph] table {data/contspill_t30.csv};
      \addplot[cl4-2]       table {data/contspill_t32.csv};
      \addplot[cl4-3]       table {data/contspill_t34.csv};
      \addplot[cl4-4]       table {data/contspill_t36.csv};

      \addlegendentry{$t=\SI{30}{\second}$}
      \addlegendentry{$t=\SI{32}{\second}$}
      \addlegendentry{$t=\SI{34}{\second}$}
      \addlegendentry{$t=\SI{36}{\second}$}

      \pgfplotsset{table/y = h_top}
      \addplot[cl4-1, emph] table {data/contspill_t30.csv};
      \addplot[cl4-2]       table {data/contspill_t32.csv};
      \addplot[cl4-3]       table {data/contspill_t34.csv};
      \addplot[cl4-4]       table {data/contspill_t36.csv};
    \end{axis}
    \end{tikzpicture}
    \caption{After the spill ends.}
    \label{fig:contspill_single_h_b}
  \end{subfigure}
  \caption{Spill height as a function of distance from spill source.
    The blue dashed line indicates the constant-spill-rate steady state.
  One may note in (a) that the height profiles reach the steady state behind the shock front.}
  \label{fig:contspill_single_h}
\end{figure}
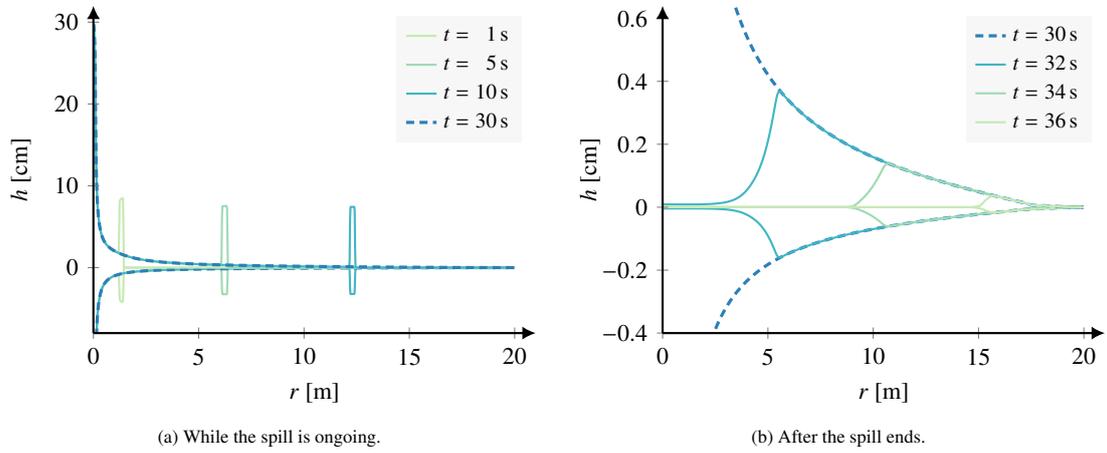

\Cref{fig:contspill_single_h} shows the height profiles $h$ of the LNG at various times $t$.
The upper lines indicate the LNG--air interface, whereas the bottom lines indicate the LNG--water interface.
In \cref{fig:contspill_single_h_a}, one can see an initial shock travelling at a velocity of about \SI{1}{\meter/\second}, which is consistent with a Froude number for the leading edge of $\sqrt 2$.
This follows naturally from the present formulation of the SWE~\cref{eq:swe}, as discussed in \citep{fyhn2019},
At $t=\SI{30}{\second}$, the pool is a monotonically decreasing function of $r$ until it vanishes at about $r = \SI{20}{\meter}$ due to evaporation.
Once the spill rate is reduced to zero, the LNG thickness starts to reduce at the origin and outwards, as shown in \cref{fig:contspill_single_h_b}.

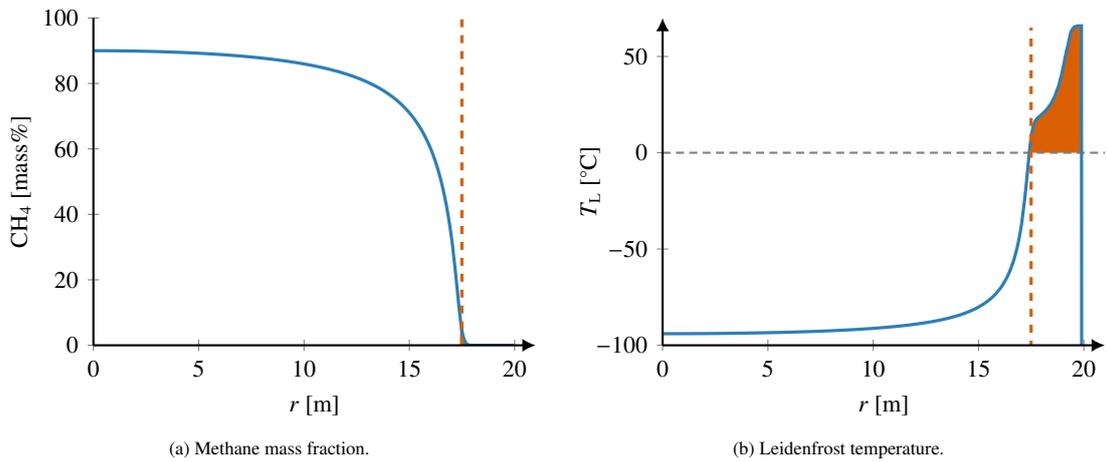
\begin{figure}[tbp]
  \centering
  \begin{subfigure}[t]{0.45\columnwidth}
    \centering
    \tikzsetnextfilename{contspill-single-c1}
    \begin{tikzpicture}
    \begin{axis}[
        xlabel={$r$ [\si{\meter}]},
        ylabel={\ce{CH4} [\si{\masspercent}]},
        xmin=0,
        xmax=21,
        ymin=0,
        ymax=100,
        y axis line style=-,
        table/y expr = 100*\thisrow{c1_frac},
        cl4-1/.append style={very thick},
      ]
      \path[name path=axis] (axis cs:0,0) -- (axis cs:21,0);

      \addplot[name path=c1, cl4-1] table {data/contspill_t30.csv};

      \addplot[color=Dark2-B, fill=Dark2-B] fill between [
        of=c1 and axis,
        soft clip={domain=17.41:19.85},
        ];

      \draw[very thick, dashed, Dark2-B]
        (\predictedRadius, 0) -- (\predictedRadius, 100);
    \end{axis}
    \end{tikzpicture}
    \caption{Methane mass fraction.}
    \label{fig:contspill_single_Tleid_a}
  \end{subfigure}
  \begin{subfigure}[t]{0.45\columnwidth}
    \centering
    \tikzsetnextfilename{contspill-single-tleid}
    \begin{tikzpicture}
    \begin{axis}[
        xlabel={$r$ [\si{\meter}]},
        ylabel={$T_\leid$ [\si{\celsius}]},
        xmin=0,
        xmax=21,
        ymin=-100,
        ymax=70,
        table/y = TleidC,
        cl4-1/.append style={very thick},
      ]
      \addplot[name path=tw, densely dashed, gray] coordinates {
        (0,0)
        (\pgfkeysvalueof{/pgfplots/xmax},0)
      };

      \addplot[name path=tleid, cl4-1] table {data/contspill_t30.csv};

      \addplot[color=Dark2-B, fill=Dark2-B] fill between [
        of=tleid and tw,
        soft clip={domain=17.41:19.85},
        ];

      \draw[very thick, dashed, Dark2-B]
        (\predictedRadius, -110) -- (\predictedRadius, 65);
    \end{axis}
    \end{tikzpicture}
    \caption{Leidenfrost temperature.}
    \label{fig:contspill_single_Tleid_b}
  \end{subfigure}
  \caption{The local (a) methane fraction and (b) Leidenfrost temperature as a function of distance from spill source for $t=\SI{30}{\second}$.
    The orange dashed line indicates the predicted $r_\RPT$ from \cref{eq:contspill-r_rpt}.
  In (b), the gray dashed line indicates $T_\water = \SI{0}{\celsius}$.}
  \label{fig:contspill_single_Tleid}
\end{figure}

In the simulations, we track the flow of each component.
This is illustrated in \cref{fig:contspill_single_Tleid_a}, where we show the radial profile of the methane fraction at the steady state $t = \SI{30}{\second}$.
The corresponding profile of the Leidenfrost temperature $T_\leid$ is shown in \cref{fig:contspill_single_Tleid_b}.
As discussed earlier, triggering of RPT may occur when $T_\leid > T_\water$.
This is highlighted in \cref{fig:contspill_single_Tleid_b}.
In both of the plots in \cref{fig:contspill_single_Tleid}, one can see that the analytically estimated $r_\RPT$ from \cref{eq:contspill-r_rpt} (orange dashed line) is very close to the simulated $r_\RPT$, which is located where $T_\leid = T_\water$.
It is interesting to note that the regions where the Leidenfrost temperature is above the water temperature coincide with regions where the thickness is small, cf.~\cref{fig:contspill_single_h_b}.

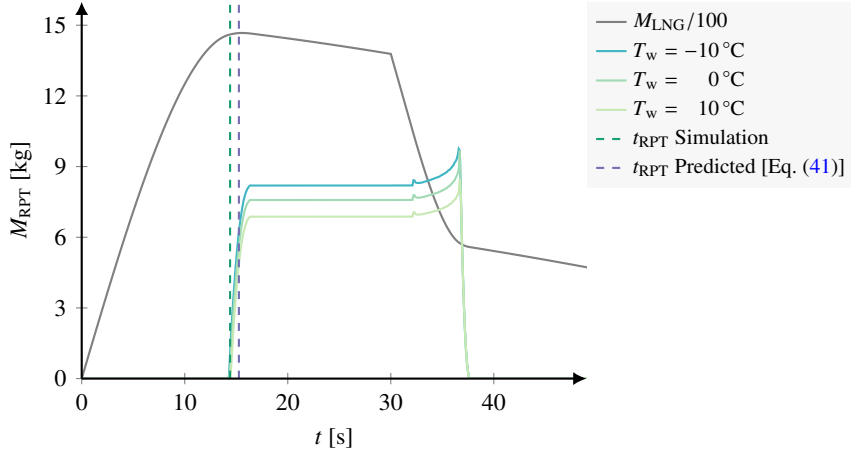
\begin{figure}[tbp]
  \centering
  \tikzsetnextfilename{contspill-single-mtrig}
  \begin{tikzpicture}
  \begin{axis}[
      height=0.4\textwidth,
      width=0.5\textwidth,
      xlabel={$t$ [\si{\second}]},
      ylabel={$M_\RPT$ [\si{\kilogram}]},
      ylabel style={at={(-0.08,0.5)}},
      xmin=0,
      xmax=49,
      ymin=0,
      ymax=16,
      ytick={0, 3, ..., 15},
      legend style={
        at={(1, 1)},
        anchor=north west,
      },
      table/x = t,
    ]
    \addplot[gray]  table[y expr = \thisrow{Mtot}/100] {data/contspill_timeseries.csv};
    \addplot[cl3-1] table[y = Mtrig_low]  {data/contspill_timeseries.csv};
    \addplot[cl3-2] table[y = Mtrig]      {data/contspill_timeseries.csv};
    \addplot[cl3-3] table[y = Mtrig_high] {data/contspill_timeseries.csv};
    \addplot[dashed, Dark2-A] coordinates {(14.4, 0) (14.4, 16)};
    \addplot[dashed, Dark2-C] coordinates {
      (\predictedTime, 0)
      (\predictedTime, 16)
    };

    \addlegendentry{$M_\LNG/100$}
    \addlegendentry{$T_\water = \SI{-10}{\celsius}$}
    \addlegendentry{$T_\water = \phantom{-1}\SI{0}{\celsius}$}
    \addlegendentry{$T_\water = \phantom{-}\SI{10}{\celsius}$}
    \addlegendentry{$t_\RPT$ Simulation}
    \addlegendentry{$t_\RPT$ Predicted [\cref{eq:contspill-t_rpt}]}
  \end{axis}
  \end{tikzpicture}
  \caption{The mass of LNG at risk of RPT for different water temperatures~$T_\water$.
    This shows that the onset of RPT is not very sensitive to the water temperature.
    A fraction of the total mass of LNG within the simulation domain is shown as a reference (gray line).
  The analytically estimated $t_\RPT$ from \cref{eq:contspill-t_rpt} is indicated by the dashed lines.}
  \label{fig:contspill_single_trigmass}
\end{figure}

As discussed in \cref{sub:triggering}, we use the Leidenfrost temperature to estimate the region where an RPT may occur.
We estimate the total mass of LNG that is in risk of triggering~$M_\RPT$ by numerically integrating the total mass per area over the area where $T_\water > T_\leid$, cf.~\cref{eq:M_rpt}.
In \cref{fig:contspill_single_trigmass}, we compare $M_\RPT$ to a fraction of the total available mass~$M_\LNG$ as a function of time.
A factor of 1/100 is used to more easily compare $M_\RPT$ and $M_\LNG$.
When the spill stops at $t=\SI{30}{\second}$, the amount of LNG that is in risk of triggering initially remains constant.
When the thickness at the origin becomes sufficiently thin, $M_\RPT$ first increases, then quickly decreases when the inner pool fully evaporates at $t\approx \SI{38}{\second}$.
To show the sensitivity of the Leidenfrost temperature, we include estimates of $M_\RPT$ for $T_\water = \SI{\pm10}{\celsius}$.
The figure shows that the time of triggering is not very sensitive to the value of the water temperature.
Finally, \cref{fig:contspill_single_trigmass} also compares $t_\RPT$ from the simulation with the estimate from simulated \cref{eq:contspill-t_rpt}.
The simulation predicts $t_\RPT \simeq \SI{14.4}{\second}$, which is slightly lower than the estimated $t_\RPT \simeq \SI[round-mode = places, round-precision = 1]{\predictedTime}{\second}$ from \cref{eq:contspill-t_rpt}.

The above case represents a fully unconstrained spill geometry.
The spill geometry will have a strong influence on the flow pattern, and thus also on the RPT phenomenon.
To illustrate this, consider a fully constrained case where an amount $M_0$ of LNG is spilled into a closed container of water where the LNG is not allowed to spread.
The RPT triggering criterion is reached as soon as the boil-off ratio reaches $\theta$.
This means that the maximum amount of LNG that can trigger is a fraction $1-\theta$ of the inital spill (\SIrange{11}{33}{\percent} for compositions $\vect z_{\mathrm A}$--$\vect z_{\mathrm C}$ in \cref{tab:compositions}).
The thickness at the triggering time is approximately $(1-\theta)h_0$, which is generally thicker than what can be achieved in the unconstrained case where the triggering thickness is in the \si{\milli\meter} range for realistic spill rates.
The energy yield of an RPT event will scale with the thickness, and thus constrained spills have a larger blast potential than unconstrained spills.

\begin{figure}[tbp]
  \centering
  \tikzsetnextfilename{contspill-comparison}
  \begin{tikzpicture}[
      label/.style={
        fill=black!5,
        above left=1ex,
        thin,
        draw,
      },
      legend matrix/.style={
        anchor=south east,
        inner sep=0.5ex,
        row sep=0.25ex,
        column sep=1ex,
        fill=black!3,
        nodes={black, font=\footnotesize, anchor=west, inner sep=0ex},
      }
    ]

    \begin{groupplot}[
        height = 0.4*\textwidth,
        width = 0.5*\textwidth,
        xlabel={},
        ylabel={},
        group style={
          group size= 2 by 2,
          vertical sep=0.75cm,
          horizontal sep=0.5cm,
          xticklabels at=edge bottom,
          yticklabels at=edge left
        },
        plot_rrpt/.style={
          ymin=0,
          ymax=34,
          table/y = r_RPT,
        },
        plot_trpt/.style={
          ymin=0,
          ymax=21,
          table/y = t_RPT,
        },
        plot_theta/.style={
          xmin=0,
          xmax=100,
          x axis line style=-,
          domain=0:100,
          table/x = theta,
          table/x expr = 100*\thisrow{theta},
        },
        plot_S/.style={
          xmin=0,
          xmax=510,
          domain=0:500,
          table/x = S,
        },
      ]
      \nextgroupplot[
        plot_rrpt,
        plot_theta,
        ylabel={$r_\RPT$ [\si{\meter}]}]
      \addplot[cl4-1] {r_rpt(x/100, 500, 7.391311)};
      \label{pgf:theta_an_1}
      \addplot[cl4-2] {r_rpt(x/100, 250, 7.391311)};
      \label{pgf:theta_an_2}
      \addplot[cl4-3] {r_rpt(x/100, 100, 7.391311)};
      \label{pgf:theta_an_3}
      \addplot[cl4-4] {r_rpt(x/100, 10, 7.391311)};
      \label{pgf:theta_an_4}

      \addplot[cl4-1, sim] table {data/contspill_S500.csv};
      \label{pgf:theta_sim_1}
      \addplot[cl4-2, sim] table {data/contspill_S250.csv};
      \label{pgf:theta_sim_2}
      \addplot[cl4-3, sim] table {data/contspill_S100.csv};
      \label{pgf:theta_sim_3}
      \addplot[cl4-4, sim] table {data/contspill_S010.csv};
      \label{pgf:theta_sim_4}

      \nextgroupplot[plot_rrpt, plot_S]
      \addplot[cl3-1] {r_rpt(0.891, x, 7.391311)};
      \label{pgf:s_an_1}
      \addplot[cl3-2] {r_rpt(0.781, x, 7.391311)};
      \label{pgf:s_an_2}
      \addplot[cl3-3] {r_rpt(0.672, x, 7.391311)};
      \label{pgf:s_an_3}

      \addplot[cl3-1, sim] table {data/contspill_m0.csv};
      \label{pgf:s_sim_1}
      \addplot[cl3-2, sim] table {data/contspill_m1.csv};
      \label{pgf:s_sim_2}
      \addplot[cl3-3, sim] table {data/contspill_m2.csv};
      \label{pgf:s_sim_3}

      \nextgroupplot[
        plot_trpt,
        plot_theta,
        xlabel={$\theta$ [\si{\percent}]},
        ylabel={$t_\RPT$ [\si{\second}]}]
      \addplot[cl4-1] {t_rpt(x/100, 500, 7.391311, 437.0, 0.1)};
      \addplot[cl4-2] {t_rpt(x/100, 250, 7.391311, 437.0, 0.1)};
      \addplot[cl4-3] {t_rpt(x/100, 100, 7.391311, 437.0, 0.1)};
      \addplot[cl4-4] {t_rpt(x/100,  10, 7.391311, 437.0, 0.1)};

      \addplot[cl4-1, sim] table {data/contspill_S500.csv};
      \addplot[cl4-2, sim] table {data/contspill_S250.csv};
      \addplot[cl4-3, sim] table {data/contspill_S100.csv};
      \addplot[cl4-4, sim] table {data/contspill_S010.csv};

      \nextgroupplot[
        plot_trpt,
        plot_S,
        xlabel={$S$ [\si{\kilo\gram\per\second}]}]
      \addplot[cl3-1] {t_rpt(0.891, x, 7.391311, 437, 0.1)};
      \addplot[cl3-2] {t_rpt(0.781, x, 7.391311, 437, 0.1)};
      \addplot[cl3-3] {t_rpt(0.672, x, 7.391311, 437, 0.1)};

      \addplot[cl3-1, sim] table {data/contspill_m0.csv};
      \addplot[cl3-2, sim] table {data/contspill_m1.csv};
      \addplot[cl3-3, sim] table {data/contspill_m2.csv};
    \end{groupplot}

    \matrix[legend matrix] at ([yshift=2ex] group c1r1.north east) {
      \node{\ref{pgf:theta_an_1}};
        & \node{\ref{pgf:theta_sim_1}};
        & \node{\quad$S = \SI{500}{\kilogram\per\second}$}; \\
      \node{\ref{pgf:theta_an_2}};
        & \node{\ref{pgf:theta_sim_2}};
        & \node{\quad$S = \SI{250}{\kilogram\per\second}$}; \\
      \node{\ref{pgf:theta_an_3}};
        & \node{\ref{pgf:theta_sim_3}};
        & \node{\quad$S = \SI{100}{\kilogram\per\second}$}; \\
      \node{\ref{pgf:theta_an_4}};
        & \node{\ref{pgf:theta_sim_4}};
        & \node{\quad$S = \phantom{5}\SI{10}{\kilogram\per\second}$}; \\
    };

    \matrix[legend matrix] at ([yshift=2ex] group c2r1.north east) {
      \node{\ref{pgf:s_an_1}};
        & \node{\ref{pgf:s_sim_1}};
        & \node{\quad$\vect z_\text{A}$, $\theta = \SI{89.1}{\percent}$}; \\
      \node{\ref{pgf:s_an_2}};
        & \node{\ref{pgf:s_sim_2}};
        & \node{\quad$\vect z_\text{B}$, $\theta = \SI{78.1}{\percent}$}; \\
      \node{\ref{pgf:s_an_3}};
        & \node{\ref{pgf:s_sim_3}};
        & \node{\quad$\vect z_\text{C}$, $\theta = \SI{67.2}{\percent}$}; \\
    };
  \end{tikzpicture}
  \caption{Comparison of $r_\RPT$ and $t_\RPT$ between analytic expression (solid lines) and simulations (filled circles) for varying (a) spill compositions and (b) spill rates.}
  \label{fig:contspill_multi_comparison}
\end{figure}
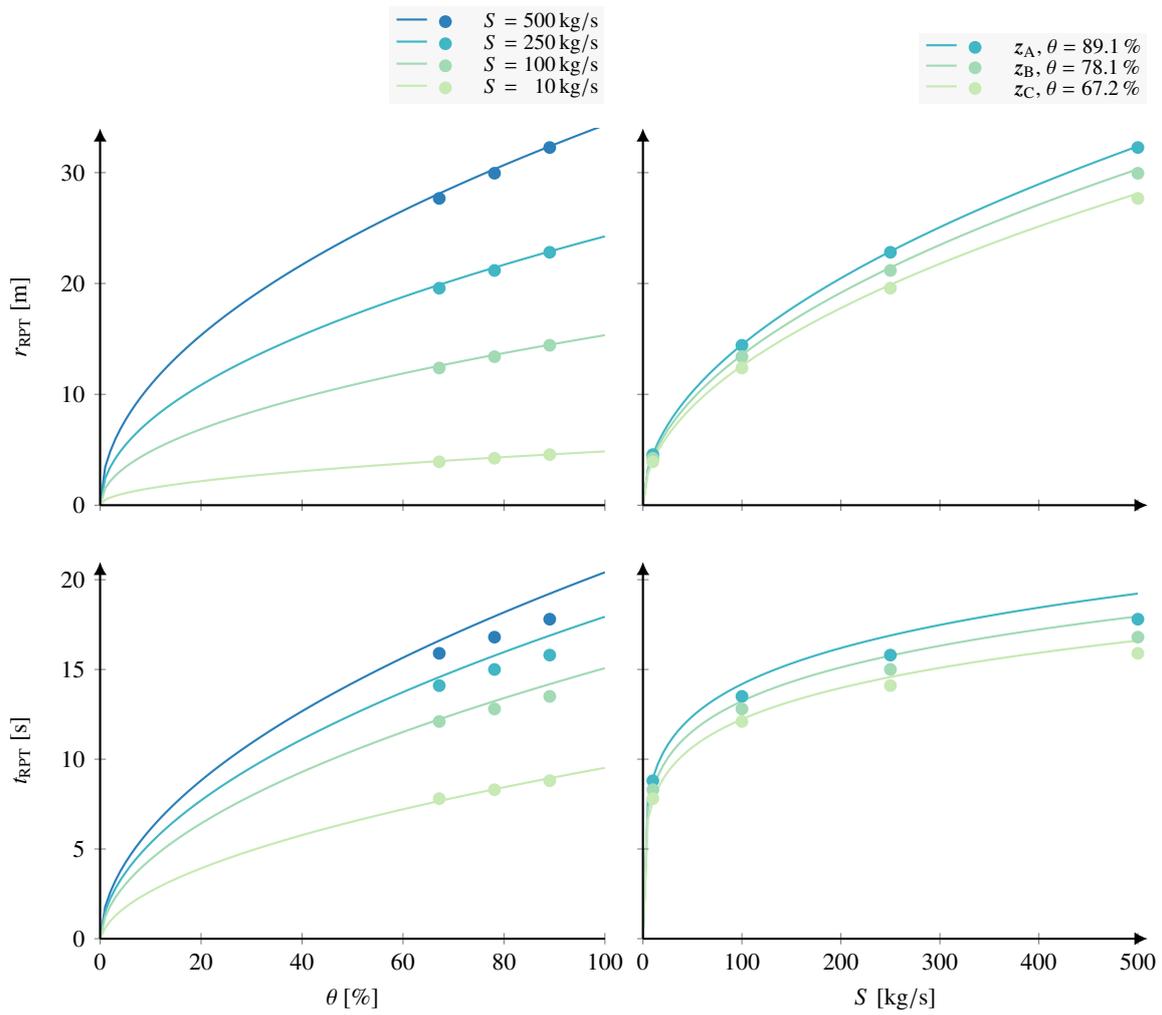

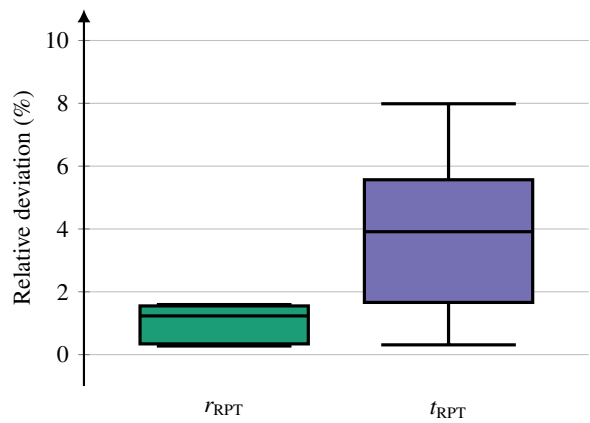
\begin{figure}[tbp]
  \centering
  \tikzsetnextfilename{contspill-comparison-errors}
  \begin{tikzpicture}
  \begin{axis}[
      height=0.4\textwidth,
      width=0.5\textwidth,
      boxplot/draw direction=y,
      boxplot/box extend=0.6,
      x axis line style={draw=none},
      x tick style={draw=none},
      enlarge y limits,
      ymajorgrids,
      set layers=standard,
      ylabel={Relative deviation (\si{\percent})},
      ylabel style={at={(-0.08,0.5)}},
      xtick={1,1.8},
      xticklabels={$r_\RPT$, $t_\RPT$},
      xmin=0.5,
      xmax=2.3,
      ymin=0,
      ymax=10,
    ]

    \addplot[very thick, fill=Dark2-A,
      boxplot prepared={
        lower whisker=0.2739,
        lower quartile=0.3414,
        median=1.2318,
        upper quartile=1.5500,
        upper whisker=1.5936,
      },
      ] coordinates {};

    \addplot[very thick, fill=Dark2-C,
      boxplot prepared={
        draw position=1.8,
        lower whisker=0.3106,
        lower quartile=1.6616,
        median=3.9129,
        upper quartile=5.5668,
        upper whisker=7.9861,
      },
      ] coordinates {};
  \end{axis}
  \end{tikzpicture}
  \caption{Boxplots of the relative deviation of analytically predicted and simulation predicted $r_\RPT$ and $t_\RPT$.}
  \label{fig:contspill_multi_errors}
\end{figure}

The earlier results indicate that the analytical estimates of $r_\RPT$ and $t_\RPT$ from \cref{eq:contspill-r_rpt,eq:contspill-t_rpt} are good approximations to the corresponding predictions from simulations.
To further validate this, we compared results for a set of simulations with different mass compositions and spill rates.
We considered the three different mass compositions listed in \cref{tab:compositions} and four different spill rates $S \in \{10, 100, 250, 500\}$ (\si{\kilogram\per\second}).
The remaining simulation parameters were similar to the above case description.
As showed in the top row of \cref{fig:contspill_multi_comparison}, the analytical estimate of $r_\RPT$ matches the simulated predictions very well.
The relative deviation is lower than \SI{2}{\percent} in every prediction, as seen in \cref{fig:contspill_multi_errors}.
The analytic estimate of $t_\RPT$ also matches well with a deviation within \SI{8}{\percent}.
The prediction of $r_\RPT$ relies only on the assumption that mass does not accumulate, while the prediction of $t_\RPT$ also assumes that the velocity and height distributions close to the shock are accurately approximated by the steady state solution.
It is therefore not unexpected that the prediction of $t_\RPT$ is less accurate than the prediction of $r_\RPT$.

\clearpage
\section{Conclusions}
\label{sec:conclusion}
We have presented a coupled model for predicting delayed RPT from an LNG spill onto water.
The model combines the shallow-water equations for capturing the spreading of LNG on water, a film-boiling heat-flux model, a fundamental triggering condition for RPT, and an extended corresponding-states EoS for thermodynamic calculations.
It was used to study an axisymmetric continuous spill case with a constant spill rate.
The results give insight into how a continuous spill scenario develops into a steady flow.
It further highlights how evaporation alters the LNG composition during the spill.
This in turn allows prediction of the regions at risk of a delayed RPT.
We also consider the influence of the spreading geometry on an RPT event.
In particular, the potential energy yield in a constrained-geometry RPT event is larger than in an unconstrained geometry.

We have also analysed a continuous tank spill in more detail based on the governing equations.
We derived a general methodology to predict the position and time of delayed RPT that is applicable to different geometries.
The method requires a steady-state solution of the tank spill and should be easy to implement with numerical codes as a post-processing analysis method.
We applied the methodology to derive simple, predictive equations for the triggering time and position of a two-dimensional axisymmetric spill.

Subsequently, we compared results from numerical simulations with the analytical predictions.
Notably, the analytical prediction of the RPT location is accurate to within \SI{2}{\percent} of that predicted from the numerical simulations.
The analytical prediction of the time of RPT is within \SI{8}{\percent} of the numerical prediction.

Unfortunately, there is a lack of relevant experimental data with which to compare the predictions.
In future work, it would be very useful to perform experiments on a similar continuous spill case in order to assess and validate both the simplified predictive equations and the simulation model.
This would also make it possible to suggest a semi-empiric form of the prediction equations for the RPT event that is based on the present analysis.
It would also be interesting to further develop predictive equations for other types of spill scenarios and other cryogenic fluids.

\section*{Acknowledgment}
This work was undertaken as part of the research project ``Predicting the risk of rapid phase-transition events in LNG spills (Predict-RPT)'', and the authors would like to acknowledge the financial support of the Research Council of Norway under the MAROFF programme (Grant 244076/O80).

\appendix
% Fakesection: References
% \bibliographystyle{elsarticle-harv}
% \bibliography{references}

% Fakesection: Nomenclature
% Define databases
\DTLnewdb{abbreviations}
\DTLnewdb{symbols}

% Add entries
% abbrevations
\addabbr{LNG}{Liquified natural gas}
\addabbr{LOC}{Loss of containment}
\addabbr{RPT}{Rapid phase transition}
\addabbr{SWE}{Shallow-water equations}
\addabbr{\eos}{Equation of State}

% symbols
\addsymb{$g_\eff$}{Effective gravitational acceleration}
\addsymb{$h$}{LNG thickness (vapor and liquid)}
\addsymb{$h_\vap$}{LNG thickness (vapor only)}
\addsymb{$M_\LNG$}{LNG mass (in total)}
\addsymb{$M_\RPT$}{LNG mass (within the RPT risk region)}
\addsymb{$\vect{m}$}{LNG mass per unit area (component vector)}
\addsymb{$m_i$}{LNG mass per unit area (component $i$)}
\addsymb{$p_\atm$}{Atmospheric pressure}
\addsymb{$\dot q$}{Heat flux from water to LNG}
\addsymb{$r_\RPT$}{Distance an LNG control mass has to travel before a possible RPT event}
\addsymb{$S$}{LNG spill rate}
\addsymb{$T_\leid$}{LNG Leidenfrost temperature}
\addsymb{$T_\shl$}{LNG superheat limit temperature}
\addsymb{$T_\bub$}{LNG bubble point temperature}
\addsymb{$T_\water$}{Water temperature}
\addsymb{$t_\RPT$}{Time an LNG control mass has to travel before a possible RPT event}
\addsymb{$\vect u$}{Horizontal velocity (LNG liquid flow)}
\addsymb{$u_B$}{Vertical velocity (LNG vapor bubbles)}
\addsymb{$z_i$}{LNG mass fraction (component $i$)}
\addsymb{$\Gamma$}{LNG boil-off fraction (general quantity)}
\addsymb{$\theta$}{LNG boil-off limit (threshold for RPT)}
\addsymb{$\Delta H^\evap$}{LNG specific enthalpy of evaporation}
\addsymb{$\delta$}{LNG buoyancy factor}
\addsymb{$\rho$}{LNG density (vapor and liquid)}
\addsymb{$\rho_\vap$}{LNG density (vapor only)}
\addsymb{$\si{\masspercent}$}{Mass percentage}

% Sort the database
%\DTLsort*{Abbreviation}{abbreviations}
%\DTLsort*{Symbol}{symbols}

% Display the contents of the database
\clearpage
\section*{Nomenclature}
\subsection*{Abbreviations}
\begin{nomenclature}
\DTLforeach*{abbreviations}{\thisAbbr=Abbreviation,\thisDesc=Description}%
  {\item[\thisAbbr] \thisDesc}%
\end{nomenclature}
\subsection*{Symbols}
\begin{nomenclature}
\DTLforeach*{symbols}{\thisSymb=Symbol,\thisDesc=Description}%
  {\item[\thisSymb] \thisDesc}%
\end{nomenclature}

\end{document}